\documentclass[fleqn,usenatbib]{mnras}

\usepackage[T1]{fontenc}
\usepackage{ae,aecompl}

\usepackage{epsf,palatino,url,graphicx,wrapfig,array,setspace,amsmath,amssymb,fancyhdr,multirow,lscape,appendix,rotating,txfonts}
\usepackage{graphics,epsfig}
\bibpunct{(}{)}{;}{a}{}{,}

\usepackage{longtable}
\usepackage{xcolor}
\usepackage{color}
\usepackage{lipsum}
\usepackage{textcomp}
\usepackage{threeparttable}
\usepackage{threeparttablex}
\usepackage{enumerate}
\usepackage{xspace}
\usepackage{hyperref}
\usepackage{color,soul}

\usepackage{pdflscape}	

\newcommand{\ka}{K${\alpha}$~}

\newcommand{\ltsima}{$\buildrel < \over \sim$}
\newcommand{\lsim}{\lower.5ex\hbox{\ltsima}}
\newcommand{\gtsima}{$\buildrel > \over \sim$}
\newcommand{\gsim}{\lower.5ex\hbox{\gtsima}}

\newcommand{\xmm}{{\it XMM-Newton}\xspace}
\newcommand{\nus}{{\it NuSTAR}\xspace}

\newcommand{\cxo}{\hbox{Chandra}\xspace}

\title[Ultra compact X-ray binaries III]{The chemical composition of the accretion disk and donor star in Ultra Compact X-ray Binaries: A comprehensive X-ray analysis }

\author[Koliopanos et al.]{
Filippos Koliopanos$^{1,2}$\thanks{E-mail: fkoliopanos@irap.omp.eu},
Mathias P{\'e}ault$^{1,2}$, Georgios Vasilopoulos$^{3}$ and Natalie Webb$^{1,2}$.
\\
$^{1}$CNRS, IRAP, 9 Av. colonel Roche, BP 44346, F-31028 Toulouse cedex 4, France\\
$^{2}$Universit{\'e} de Toulouse; UPS-OMP; IRAP, Toulouse, France \\
$^{3}$Department of Astronomy, Yale University, PO Box 208101, New Haven, CT 06520-8101, USA\\
}

\date{Accepted XXX. Received YYY; in original form ZZZ}

\pubyear{2019}

\begin{document}
\label{firstpage}
\pagerange{\pageref{firstpage}--\pageref{lastpage}}
\maketitle

\begin{abstract}
We have analyzed the X-ray spectra of all known Ultra Compact X-ray Binaries (UCXBs), with the purpose of constraining the chemical composition of their accretion disk and donor star. Our investigation was focused on the presence (or absence) of the Fe \ka emission line, which was used as the probe of chemical composition of the disk, based on previously established theoretical predictions for the reflection of X-ray radiation off the surface of C/O-rich or He-rich accretion disks in UCXBs. We have contrasted the results of our spectral analysis to the history of type I X-ray bursts from these systems, which can also indicate donor star composition.  We found that UCXBs with prominent and persistent iron \ka emission also featured repeat bursting activity. On the other hand, the UCXBs for which no iron line was detected, appear to have few or no type I X-ray bursts detected over more than a decade of monitoring. Based on Monte Carlo simulations, demonstrating a strong correlation between the Fe \ka line strength and the abundance of C and O in the accretion disk material and given the expected correlation between the H/He abundance and the recurrence rate of type I X-ray bursts, we propose that there is a considerable likelihood that UCXBs with persistent iron emission have He-rich donors, while those that do not, likely have C/O or O/Ne/Mg-rich donors. Our result strongly advocate for the development of more sophisticated simulations of X-ray reflection from hydrogen-poor accretion disks.
 \end{abstract}




\section{Introduction}

Ultracompact X-ray binaries (UCXBs) are X-ray binary systems (XRBs) defined by their extremely short orbital periods of less than ${\sim}$ one hour. The orbital size defined by such short periods is too small to encompass the main sequence or red giant stars that make up for the donor star in typical low-mass XRBs (LMXBS), (e.g. \citealt{1984ApJ...283..232R}; \citealt*{1986ApJ...311..226N}). UCXBs consist of a neutron star (although black holes have not been ruled out), that is accreting matter from degenerate companion star, a Roche-lobe filling white dwarf  or helium star  \citep*[e.g.][]{1993ARep...37..411T, 1995ApJS..100..233I, 1995xrbi.nasa..457V, 2003ApJ...598.1217D, 2005ApJ...624..934D}.
 
UCXB formation history can be divided into three major schemes. White dwarf (WD) neutron star (NS) binaries that gradually lose angular momentum due to gravitational radiation until the WD fills its Roche lobe and starts losing mass to the NS \cite{1975MNRAS.172..493P}. Semi-attached binaries reaching an exceptionally short orbital periods, resulting in high mass accretion rates from hydrogen-depleted non-degenerate donors \cite{1986ApJ...304..231N} or -- in case of a companion star with a helium burning core at the time of contact -- a helium star donor \cite{1986A&A...155...51S}. Therefore, depending on the evolutionary track that leads to their creation UCXBs can have a variety of different donors ranging from Helium stars to C/O or O/Ne/Mg WDs \citep*[e.g.][]{1986A&A...155...51S,2002ApJ...565.1107P,2002A&A...388..546Y, 2004ApJ...607L.119B}.

UCXBs offer the unique opportunity to study accretion of matter with extraordinary chemical composition. Furthermore, they are  are unique laboratories for the study of binary evolution theory and particularly the common-envelope phase and its relevance to type Ia supernovae and are strong gravitational wave sources in the low-frequency regime where the ESA/NASA LISA will be sensitive \citep[e.g.,][]{2010NewAR..54...87N,2017arXiv170200786A,2017JPhCS.840a2024C}. They stand as excellent candidates for in-depth multi-messenger studies of accretion onto compact sources. As population studies and disk stability and evolution analyses predict specific ratios of He-rich vs C/O - O/Ne/Mg  systems, a key step in understanding of UCXB formation and evolution is the determination of the nature of the donor star composition and whether it lies in the one or the other category \citep[][]{2004ApJ...603..690B,2010MNRAS.401.1347N,2013ApJ...768..184H}.

The most straight forward approach to the determination of donor star composition would the spectral analysis of optical emission originating in the accretion disk and donor star. Indeed, for a fraction of UCXBs indications of extra-solar elemental abundances have been presented using optical and UV spectroscopy. \cite{2004MNRAS.348L...7N, 2006A&A...450..725W} and \cite{2006MNRAS.370..255N} detected strong C and O lines in the optical spectra of two UCXBs (4U~1626-67, 4U 0614+09) which also lacked any evidence for the presence of H or He, suggesting the two source may host C/O rich donors. Furthermore, \cite{2004MNRAS.348L...7N} argued that 4U~1543-624 \& 2S~0918-549 show optical spectra similar to 4U 0614+09. On the other hand, a He/N ratio consistent with a He-rich donor was indicated for 4U~1916-05 \citep{2006MNRAS.370..255N}. While optical spectroscopy can provide good indications for the chemical composition of UCXB donor stars, due to their small sized accretion disks \citep{1994A&A...290..133V} the optical emission from UCXBs is quite faint, with V-band absolute magnitudes typically larger than $\approx$5 and for distances ranging from $\approx$3-12\,kpc \citep[e.g.][]{2004MNRAS.348L...7N, 2006MNRAS.370..255N}. 
Therefore, establishing definitive evidence of the donor star composition -- using optical spectroscopy -- is a challenging task that can only be attempted using the latest generation of $>$8\,m telescopes.
 
As their name implies UCXBs are powerful sources of X-ray emission, the result of mass accretion at high mass accretion rates due to Roche lobe overflow of the donor star \citep[see review by][]{2010NewAR..54...87N}. Furthermore, the rate and stability of mass accretion depends strongly on the donor mass and chemical composition and therefore on the evolutionary track that for UCXB formation \citep[e.g.][and references therein]{2013ApJ...768..184H}. This paper is focused on the X-ray emission of UCXBs and demonstrates how X-ray spectroscopy, combined with numerical simulations of fundamental radiative processes, in the accretion disk and its vicinity can provide credible indications on the chemical composition of the accretion disks and donor stars of UCXB systems. In Section~\ref{sec:X-ray_diag} we lay out the main principles of X-ray diagnostics of the chemical composition of UCXBs, followed by Section~\ref{sec:analysis}, where we present the results of X-ray spectroscopic analysis of all known UCBXs. We discuss our findings and present our conclusions in Sections~\ref{sec:discu} and \ref{sec:conc}. 

\section{X-ray diagnostics of UCXBs}
\label{sec:X-ray_diag}

The primary X-ray emission originates in the inner accretion disk and/or the surface layer hot optically thick matter on the NS, known as the boundary layer \citep[][]{2001ApJ...547..355P}. A secondary emission -- evident by the hard X-ray emission with a non-thermal spectrum -- originates in a corona of hot thermal electrons in the vicinity of the disk and central engine. The hot electrons up-scatter seed photons from the disk and boundary layer, producing emission with a spectrum characteristic of unsaturated thermal comptonization \citep[e.g.][and references therein]{2010LNP...794...17G}. A fraction of these hard photons may be "reflected" (partially absorbed and re-emitted and/or scattered) off the accretion disk surface producing an additional emission component known as X-ray reflection \citep{1974A&A....31..249B,1988ApJ...331..939W,1988ApJ...335...57L, 1991MNRAS.249..352G}. 

X-ray reflection spectra  are enriched with multiple emission lines -- the result of K-shell emission from heavy elements present in the accreting material. However, most of the line photons are heavily diluted by the primary continuum photons (registered together with the reflected emission). Due to the high fluorescence yield and chemical abundance of iron,  a bright Fe  \ka emission line at $\approx 6.4-6.9$\,keV with an equivalent width (EW) typically of the order of $\approx100$\,eV  \citep[e.g.][]{2010ApJ...720..205C} is usually detected in the spectra of normal LMXBs with main sequence or red giant 
donors. Spectral analysis of the reflection component -- and particularly Fe \ka line -- can provide indications for the disk chemical composition and its physical parameters such as temperature and rotational velocity and size \citep[e.g.][]{1989MNRAS.238..729F,1993MNRAS.262..179M,1997MNRAS.289..175M,2001MNRAS.327...10B, 2010ApJ...718..695G,2013ApJ...768..146G, 2013MNRAS.432.1264K}.

In the case of UCXBs, X-ray reflection can -- in principle -- provide valuable insight into the abundances of chemical elements in the disk and donor star, thus revealing its nature.
More specifically, the presence of oxygen and neon emission features --
that appear in the spectra of reprocessed emission from the accretion disk and white dwarf surface --  \cite[e.g.][]{2010MNRAS.407L..11M} and K-edges stemming from
absorbing material in the vicinity of the disk \cite[e.g.][]{2010ApJ...725.2417S}
could provide direct indication of a C/O or O/Ne-rich disk and donor star.
However, due to increased interstellar absorption below 1\,keV and 
contamination of the reflected component by the primary emission, detection of these features with sufficient accuracy, often proves to be difficult. 

In \cite{2013MNRAS.432.1264K} we used the Monte Carlo method to develop -- for the first time -- simulations of reflection from C/O or O/Ne/Mg, hydrogen poor disks. {Perhaps paradoxically, our study demonstrated that the most striking effect of the hydrogen poor, C/O rich material is not an appearance of strong fluorescent lines of carbon and oxygen -- as one might expect -- but nearly complete disappearance of the iron \ka line. This is caused by the “screening” of iron by the much more abundant carbon and oxygen. In optically thick,  neutral material with solar abundances, the most likely process for incident photons with h$\nu{\gtrsim}$7.1\,keV (Fe K-shell photo-ionisation threshold) is absorption by iron due to the photo ionisation of its atoms. The process is followed by the emission of fluorescent Fe \ka photons at 6.4\,keV \citep[with a fluorescence yield of approximately one third][]{1972RvMP...44..716B}. Consequently, the majority of photons with energies above this threshold will be absorbed by iron and will, therefore, contribute to its fluorescent line.}

{In the case of a C/O (or O/Ne) white dwarf though, the extraordinary overabundance of oxygen makes it the dominant absorbing agent even at energies well beyond its own K-edge (see Fig.~1 in \citealt{2013MNRAS.432.1264K}). As a result only a very small fraction of incident photons will fuel the iron \ka line. Despite the fact that the O \ka line produced in this environment is boosted significantly, it is still strongly diluted by the primary continuum and therefore is difficult to detect. A much more visible effect is the significant attenuation or complete disappearance of the iron line.
Helium, on the other hand, is not capable of screening iron, due to its smaller charge and correspondingly smaller absorption cross-section at the iron K-edge. Furthermore, helium is fully ionized throughout most of the accretion disk, for the entire range of accretion rates, observed in XRBs. Therefore in the case of a He-rich donor X-ray reflection is practically indistinguishable from the case of H-rich accretion disks of "standard" XRBs, and the iron line has its nominal strength .}

{More specifically, for a source of moderate  luminosity (${\rm L}\rm_X\lesssim$ a few $10^{37}\rm erg\,s^{-1}$) the predicted suppression of the Fe $\rm K{\alpha}$ line -- in the case of a C/O or O/Ne/Mg WD donor -- translates to a more than an order of magnitude decrease of the EW of the line. On the other hand, in the case of a He-rich donor the iron line  remains unaffected, i.e., with the EW near the typical value of
the order of${\sim}$100\,eV \citep[e.g.,][]{2010ApJ...720..205C,2010A&A...522A..96N}. These predictions were observationally tested in \cite{2014MNRAS.442.2817K} where we demonstrated that this effect can be used as a diagnostic of the chemical composition of the accretion disk and donor star of UCXBs.}

In addition to the direct approach of spectral analysis, the composition of the accreted matter can be indirectly inferred by monitoring the systems' bursting activity in NS powered UCXBs (all known UCXBs host NSs). Type I X-ray bursts are the result of accumulation of H and/or He on the NS surface until it reaches the necessary conditions to ignite as a thermonuclear flash (e.g. \citealt{1976ApJ...205L.127G}; \citealt{1975ApJ...195..735H} and for a detailed review \citealt{2006csxs.book..113S}).  Based on their recurrence rate  and temporal characteristic (duration, light-curve) one can indicate the donor type chemical composition in UCXBs, from their bursting activity. However, it must be noted that the persistence and the rate of the accretion process are also crucial factors determining the bursting activity of X-ray binaries.    

In this paper we build on the predictions of \cite{2013MNRAS.432.1264K} and the findings of \cite{2014MNRAS.442.2817K} in order to provide an assessment of the donor star type of all known UCXBs. We study X-ray spectra from various X-ray observatories for all known UCXBs and use our simulations to estimate the chemical composition of disk and donors star, based on the presence and strength of the iron \ka line. We evaluate our results by comparing with previous considerations, based on available optical analysis.  We further account for the bursting activity of each source -- based on the latest available monitoring information -- contrasting their activity to our own estimations in order to scrutinize our classification. Our aim is to use X-ray spectroscopy to diagnose the chemical composition of the donor star of UCXBs, and provide a substantial addition to the study of the evolution of double degenerate systems.

\section{Data selection, extraction \& analysis}
\label{sec:analysis}

{ We analyse all 14 confirmed UCXBs (i.e.~with a measured optical period of less than 1\,hr) plus 2 well known candidate UCXBs, albeit with tentative orbital period estimation (4U~0614+091, 4U~1728-34).} We primarily use \xmm observations since they provide the optimum combination of spectral resolution and effective area, in the 6-7\,keV, to allow for clear line detection and EW estimation but more importantly to attain a tight upper limit in sources were no line is detected. For this reason we selected observations at high count rate. In case were \xmm data were unavailable we made use of \nus or RXTE observations. 
For all UCXBs we report their bursting activity (or lack there of), based on the archive all high energy all-sky monitors, such as the All Sky Monitor (ASM, on board RXTE), BeppoSAX, JEM-X (on board INTEGRAL), Fermi-LAT and MAXI. We reference the works of \cite{2016ApJ...826..228J} and \cite[][for 4U~1820-30]{2003ApJ...595.1077C}, \cite{2013ApJ...767L..37D,2017ApJ...836..111K} and \cite{2018ApJ...858L..13S} for IGR~J17062-6143, the Multi-INstrument Burst ARchive (MINBAR\footnote{\url{http://burst.sci.monash.edu/wiki/index.php?n=MINBAR.V09ReleaseNotes}}) and the SRON catalog of X-ray bursters\footnote{\url{https://personal.sron.nl/~jeanz/bursterlist.html}} (in 't Zand et al.). In Table~\ref{tab:sources_all} we present the list of sources analysed in this work, the ObsIDs of the selected observations along with their distances and count rates. 

\begin{table*}
\caption{ Observations analyzed in this work. UCXBs are tabulated along with the count-rates of selected observations, their most recent distance and orbital period estimations and their relevant references.  }
\begin{tabular}{ lccccc }
  Source              &  Distance                       & Period            &  Telescope  & ObsID         &Net count rate     \\
                      &  (kpc)                          & (minutes)         &             &               &(cts/s)            \\
  \hline                                                                                                   
  4U~0513-40          & $12.1{\pm}0.6^a$                  & $17^b$          &  \xmm       &  0151750101   & 35.89${\pm}$0.05  \\   
  4U~0614+091         & $3.2{\pm}0.5^c$                   & $51?^d$         &  \xmm       &  0111040101   & 230.9${\pm}$0.14 \\    
  2S~0918-549         & $5.4{\pm}0.8^e$                   & $17.4^f$        &  \xmm       &  0061140101   & 55.45${\pm}$0.04  \\   
  XTE~J0929-314       & $8^{+7\ }_{-3}{^g}$               & 43.6$^h$        &  RXTE       &  70096030500  & 30.12${\pm}$0.14  \\   
  4U~1543-624         & $7.0?^i$                          & $18.2^i$        &  \xmm       &  0061140201   & 208.7${\pm}$0.07 \\   
                      &                                   &                 &  RXTE       &  P20071      & 382.42${\pm}$0.98  \\  
  4U~1626-67          & $8^{+5}_{-3}$ $^j$                & $42^j$          &  \xmm       &  0111070201   & 28.03${\pm}$0.05  \\ 
                      &                                   &                 &  \nus       &  30101029002 & 12.42${\pm}$0.08  \\  
  IGR~J17062-6143     & $7.3{\pm}0.5^{v}$                 & $38^v$          &  \xmm       &  0790780101   & 16.5${\pm}$0.02 \\    
  4U~1728-34          & $5.2{\pm}0.8^k$                   & $10.8?^k$       &  \xmm       &  0671180501   & 123.3${\pm}$0.05 \\    
  XTE~J1751-305       & $8^{+0.5\ }_{-1.3}$ $^{l}$        & $42^{m}$        &  \xmm       &  0154750301   & 80.95${\pm}$0.05  \\   
  Swift~J1756.9-2508  & $8.5{\pm}4^{n}$                   & 54.7$^{n}$      &  \xmm       &  0830190401   & 19.45${\pm}$0.09  \\  
                      &                                   &                 &  RXTE       &   P92050      & 21.51${\pm}$0.08  \\  
  XTE~J1807-294       & $8^{+4}_{-3.3}{^g}$               & $40.1^{o}$      &  \xmm       &  0157960101   & 32.50${\pm}$0.07  \\  
  4U~1820-30          & $7.9{\pm}0.4^a$                   & $11^{p}$        &  \nus       &  90401323002  & 92.11${\pm}$0.22  \\   
  4U~1850-087         & $6.9{\pm}0.3^a$                  & $20.6^{q}$      &  \xmm       &  0142330501   & 43.52${\pm}$0.07  \\  
  4U~1916-05          & $9.3{\pm}1.4^{r}$                 & $50^{s}$        &  \xmm       &  0085290301   & 52.87${\pm}$0.07  \\   
  M15~X-2             & $10.4{\pm}0.5^a$                  & $22.6^{t}$      &  \xmm       &  0087350801   & 79.73${\pm}$0.10  \\  
  NGC~6440~X-2        &  8.5${\pm}0.4^a$                  & 57.3$^{u}$      &  RXTE       &  94044040600  & 19.46${\pm}$0.012 \\   
  \hline                                                                     
  \end{tabular}           
  
 References:
 $^a$\cite{2010arXiv1012.3224H};   $^b$\cite{2009ApJ...699.1113Z}; 
 $^c$\cite{1992A&A...262L..15B};   $^d$\cite{2008PASP..120..848S}; 
 $^e$\cite{2005A&A...441..675I};   $^f$\cite{2011ApJ...729....8Z};
 $^g$\cite{2006AIPC..840...50G};   $^h$\cite{2002ApJ...576L.137G};
 $^i$\cite{2004ApJ...616L.139W};   $^j$\cite{1998ApJ...492..342C};
 $^k$\cite{Galloway10};            $^{l}$\cite{2008MNRAS.383..411P}; 
 $^{m}$\cite{2002ApJ...575L..21M}; $^{n}$\cite{2007ApJ...668L.147K}; 
 $^{o}$\cite{2003ATel..127....1M}; $^{p}$\cite{1987ApJ...312L..17S};
 $^{q}$\cite{1996MNRAS.282L..37H}; $^{r}$\cite{2008ApJS..179..360G};        
 $^{s}$\cite{1982ApJ...253L..67W}; $^{t}$\cite{2005ApJ...634L.105D};
 $^{u}$\cite{1994A&AS..108..653O}; $^{v}$\cite{2017ApJ...836..111K,2018ApJ...858L..13S}.
\label{tab:sources_all}
 \end{table*}

\subsection{\xmm spectral extraction}
\label{sec:xmm} 

For the \xmm data, we only considered the EPIC-pn detector. EPIC-pn has the largest effective area of the three CCD detectors on board the \xmm ( approximately five times higher than 
that of MOS detectors at  $\approx 7$\,keV). Furthermore, all sources have very high count-rates registering ${\gtrsim}5{\times}10^{5}$\, photons for each of the observations considered, thus providing sufficient statistics to robustly detect the presence of emission lines -- or place tight EW upper limits when not detected. Using one type of detector ensures simplicity and self-consistency in our analysis.
Therefore the following description of data analysis refers only to the EPIC-pn instrument.

The data were handled using the \xmm data analysis software SAS version 18.0.0. and the calibration files released\footnote{XMM-Newton CCF Release Note: XMM-CCF-REL-367}  on March 29, 2019. Following the standard \xmm guidelines, we filtered all observations for high background-flaring activity, by extracting high-energy light curves (10$<$E$<$12\,keV) with a 100\,s bin size. By placing appropriate threshold count rates for the high-energy photons, we filtered out time intervals that were affected by high particle background.
All \xmm observations considered for this work have been checked and for pile-up and were found to be unaffected. Only in the case of 4U~1820-30, all available \xmm observations suffered from pile-up, for this reason a recent \nus observation was used for our analysis.

With the exception of 4U~0513-40 (ObsID:0151750101) and 4U~1850-087 (ObsID:0142330201), which were taken in full window Imaging mode, all other \xmm observations had pn operating in Timing Mode. For the two sources observed in Imaging mode, spectra were extracted from a circular region with a radius 45\arcsec  centered at the core of the point spread function (psf) of each source. We thus ensured the maximum encircled energy fraction\footnote{See \xmm Users Handbook \S3.2.1.1 \\  http://xmm-tools.cosmos.esa.int/external/xmm\_user\_support/ \\ documentation/uhb/onaxisxraypsf.html} within the extraction region.  The extraction and filtering process followed the nominal guidelines provided by the \xmm Science Operations Centre (SOC\footnote{http://www.cosmos.esa.int/web/xmm-newton/sas-threads}). Spectral extraction was done with SAS task \texttt{evselect} using the standard filtering flags (\texttt{\#XMMEA\_EP \&\& PATTERN<=4} for pn), and SAS tasks \texttt{rmfgen} and \texttt{arfgen} were used to create the redistribution matrix and ancillary file,  respectively. The spectra were regrouped to have at least 25 counts per bin and analysis in order to allow for the use if ${\chi}^{2}$ statistics in the spectral fitting.
Source photons for all pn observations, taken in timing mode, were extracted from the RAW
coordinate event file. More specifically the source spectra where extracted from 12 columns comprising the core source emission located  between RAWX 25 and 50  for all sources. Background was extracted from  RAWX  3 to 5, where RAWX is the coordinate along the column axis.

\subsection{\nus spectral extraction}
\label{sec:nus} 

For the analysis of the \nus data, we used version 1.8.0 of the \nus data analysis system (\nus DAS) and instrumental calibration files from CalDB v20180312. We used the \texttt{NUPIPELINE} script to calibrate and clean the data, considering the standard settings reference in the \nus observer's manual. We reduced internal high-energy background, and screened out all passages through the South Atlantic Anomaly (settings SAACALC$=$3, TENTACLE$=$NO and SAAMODE$=$OPTIMIZED). Source and background spectra were extracted using the \texttt{NUPRODUCTS} routine and instrumental responses were then produced for each of the two focal plane CCD detectors (FPMA/B). The source spectrum was extracted from a circular region  of 50{\arcsec} radius, and the background was estimated from a region of the same size, in a blank sky region on the same detector as the source, as far from the source as possible in order to avoid contribution from the PSF wings. Standard PSF, alignment and vignetting corrections were applied upon extraction.

\subsection{RXTE spectral extraction}
\label{sec:rxte} 

We used data from the RXTE mission for sources XTE~J0929-314, Swift~J1756.9-2508 and NGC~6440~X-2, for which no high quality data were available from other observatories.
More specifically  we obtained and analyzed spectra from the Proportional Counter Array \citep[PCA,][]{2006ApJS..163..401J}  instrument on board RXTE. While PCA lacks the energy resolution of the CCD detectors on board \xmm and \nus it has a very high effective area, yields spectra with very high S/N ratio and is ideal for emission line-detection (and EW estimation), even if it cannot provide extremely accurate centroid energy or line-width estimation.

The PCA detector consists of five Proportional Counter Units (PCUs), with a combined effective area of $\sim7000\,cm^2$. Each PCU is composed of three layers of xenon (90\%) and methane (10\%) composites. The Incident photons with energies exceeding 20\,keV are mostly detected in the top layer (layer 1). We, therefore consider only spectra from this layer. Following the guidelines\footnote{https://heasarc.gsfc.nasa.gov/docs/xte/pca/doc/rmf/pcarmf-11.7/} from the RXTE guest observer facility (GOF), we ignore energy channels below 3\,keV recommendation. We also removed all channels above 30\,keV. Signal-to-noise ratio drops significantly above ${\sim}30$\,keV, and for this work we are only interested in the energy range below 30\,keV.

We filtered out observation intervals where the elevation angle was less than 10 degrees (to avoid possible Earth occultations) and also any data that may have been received during passage from the South Atlantic Anomaly. Furthermore, we excluded any data affected by electron contamination, were offset by more than 0.02 degrees or taken 150 seconds before, through 600 seconds after any PCU2 breakdowns, during the observation. Source spectra were extracted and background emission was modeled using standard routines from the {\small FTOOLS} package and the latest model for bright sources, provided by the GOF\footnote{http://heasarc.gsfc.nasa.gov/docs/xte/pca\_news.html}. 

\subsection{Spectral analysis }

\begin{table*}
 \caption{Best fit parameters for the continuum emission spectra. 4U~1626-67 is the only UCXB that is a high-B NS, X-ray pulsar and is tabulated separately.   All errors are in the 1$\sigma$ confidence range.}
 \begin{center}
\scalebox{0.8}{   \begin{tabular}{lcccccccccc}
     \hline\hline\noalign{\smallskip}
     \multicolumn{1}{c}{Source} &
          \multicolumn{1}{c}{nH} &
     \multicolumn{1}{c}{k${\rm T_{dBB}}$} &
     \multicolumn{1}{c}{${\rm R_{dBB}}^{a}$} &
     \multicolumn{1}{c}{${\rm \Gamma}$} &
     \multicolumn{1}{c}{${\rm K_{po}}^{b}$} &
     \multicolumn{1}{c}{k${\rm T_{BB}}$}&
     \multicolumn{1}{c}{${\rm R_{BB}}^{c}$} &
     \multicolumn{1}{c}{${\rm {L}^{d}}$} &
          \multicolumn{1}{r}{red. ${\chi^2/dof}$ }\\

      \multicolumn{1}{c}{} &
      \multicolumn{1}{c}{${\times}10^{22}$} &
      \multicolumn{1}{c}{keV} &
      \multicolumn{1}{c}{km} &
      \multicolumn{1}{c}{} &
      \multicolumn{1}{c}{[${\times}10^{-3}$]} & 
      \multicolumn{1}{c}{}keV &
      \multicolumn{1}{c}{km} &
      \multicolumn{1}{c}{{${\times}10^{36}${erg\,s$^{-1}$}}} &
      \multicolumn{1}{c}{} \\
      \noalign{\smallskip}\hline\noalign{\smallskip}

4U~0513-40$^{e}$    & 0.22${\pm}0.01$ & 0.11${\pm0.001}$       &  632${\pm}31.4$          & 2.01${\pm}0.01$        & 23.3${\pm}0.65$        & --                     & --              &  2.24  & 1.12/163  \\
4U~0614+091$^{e}$   & 0.50${\pm}0.01$ & 0.10${\pm0.001}$       &  3526${\pm}702$          & 2.21${\pm}0.01$        & 306${\pm}1.11$         & --                     & --              &  5.32  & 1.16/1888 \\
2S~0918-549         & 0.07${\pm}0.001$& 0.59${\pm0.01}$        &  15.3${\pm}2.28$         & 0.96$_{-0.32}^{+0.22}$ & 3.23$_{-1.74}^{2.23 }$ & 1.23${\pm}0.03$        & 0.57${\pm}0.09$ &  0.68  & 1.08/1848 \\
XTE~J0929-314       & 0.12$^{f}$      & 0.57${\pm0.12}$        &  85.1$_{-54.6}^{+126 }$  & 1.88${\pm}0.08$        & 89.5$_{-16.1}^{21.4 }$ & --                     &  --             & 3.69  & 0.48/49   \\
4U~1543-624$^{e}$   & 0.29$^{f}$      & 0.38${\pm0.12}$     &${14.2_{-0.56}^{+0.64}}^{g}$ &  --                    & --                     & 1.53${\pm}0.01$        & 0.16${\pm}0.03$ & 3.64   & 1.03/1795 \\
IGR~J17062-6143     & 0.14${\pm}0.01$ & 0.50${\pm}0.03$        & 14.5${\pm}1.04$          &1.22$_{-0.47}^{+0.35}$  & 1.28$_{-0.85}^{1.60 }$ & 1.00$_{-0.10}^{+0.08}$ & 0.49${\pm}0.02$ &  0.36  & 1.14/175 \\
4U~1728-34          & 2.25${\pm}0.01$ & 0.72${\pm}0.03$        & 10.7${\pm}1.64$          &  --                    &  --                    & 2.27${\pm}0.01$        & 0.10${\pm}0.01$ &  5.70  & 1.18/1748 \\
XTE~J1751-305       & 1.39${\pm}0.02$ & 0.90${\pm0.01}$        &  9.89$_{-0.56 }^{+1.46}$ &  --                    & --                     & 3.45${\pm}0.01$        & 0.49${\pm}0.01$ &  8.21  & 1.05/1695 \\
Swift~J1756.9-2508  & 5.47${\pm}0.03$ & 0.94${\pm0.07}$        &  10.6${\pm1.11}$         &  --                    &  --                    & 3.20$_{-0.16}^{+0.20}$ & 0.41${\pm}0.05$ &  5.05  & 1.04/1859 \\
XTE~J1807-294$^{e}$ & 0.57            & 0.59${\pm0.01}$        &  11.7$_{-3.36 }^{+5.83}$ &  --                    &  --                    & 2.73${\pm}0.10$        & 0.38${\pm}0.01$ &  1.84  & 1.01/1632 \\
4U~1820-30          & 0.14$^{f}$      & 0.67$_{-0.06}^{+0.05}$ &  192${\pm}9.74$     &${2.12_{-0.41}^{+0.57}}^{h}$ & 665$_{-32.3}^{31.8}$   & --                     & --              &  65.4   & 1.00/589 \\
4U~1850-087$^{e}$   & 0.25$^{f}$      & 0.45${\pm0.01}$        &  16.7${\pm}0.74$         &  --                    &  --                    & 1.96${\pm}0.04$        & 0.54${\pm}0.02$ &  14.4  & 1.19/161 \\
4U~1916-05          & 0.53${\pm}0.02$ & 0.11${\pm0.001}$       &  859${\pm}41.3$          & 1.93${\pm}0.03$        & 64.5$_{-1.43}^{1.46 }$ & 1.59${\pm}0.05$        & 0.59${\pm}0.12$ &  8.96  & 1.09/1797 \\
4U~2129+11 (M15~X-2)& 0.20${\pm}0.02$ & 0.10$_{-0.02}^{+0.01}$ &  354${\pm}53.2$          & 1.45${\pm}0.02$        & 44.6$_{-0.96}^{0.11 }$ &  --                    &  --             &  11.7  & 1.05/1711 \\
NGC~6440~X-2        &  0.31$^{f}$     &  --                    &   --                     & 1.85${\pm}0.03$        & 62.6$_{-4.52}^{4.84 }$ &  --                    &  --             &  7.34  & 0.48/35 \\
      \noalign{\smallskip}\hline\noalign{\smallskip}
4U~1626-67$^{e}$    & 0.12${\pm}0.01$ & 0.36${\pm0.01}$        &  95.8$_{-3.70}^{+9.15}$  & 0.72${\pm}0.01$        & 6.04$_{-0.05}^{0.06 }$ & --                     & --              &  162   & 1.04/162 \\

     \hline\hline\noalign{\smallskip}
     \end{tabular}   }
 \end{center}
  
  $^{a}$ {$R_{\rm dBB}$ (the inner radius of the accretion disk in km) is inferred from  {\texttt {diskbb}} model, by solving K$_{\rm dBB}$=${\rm(R_{\rm dBB}/{D_{10}})^{2}}\,\cos{i}$, for $R_{\rm dBB}$. `K$_{\rm BB}$' is the normalisation of the  \texttt{diskbb} model, ${\rm D_{10}}$ is distance in units of 10\,kpc and $i$ is the inclination. In all sources this was assumed to be 60$^{o}$, with the exceptions of 4U~1916-05 and  4U~0513-40 (where $i$ is set to 85$^{o}$), which are known edge-on viewed systems \citep[][]{2004A&A...418.1061B,2011MNRAS.414L..41F}. With the exception of XTE J19029-314 (for which the K estimation has considerably high $1\,{\sigma}$ error bars), errors in radius estimation are dominated by the 1\,${\sigma}$ errors in the distance estimation (see Table \ref{tab:sources_all}). }\\
  $^{b}$ {Power-law component normalisation constant: photons/keV/cm$^2$/s at 1\,keV} \\
  $^{c}$ {Size of the spherical black body component (\texttt{bbodyrad} model), estimated from K$_{\rm BB}$=${\rm {R_{\rm BB}}^{2}/{D_{10}}^2}$, where K$_{\rm BB}$ is the normalization parameter of the \texttt{bbodyrad} model.}\\
  $^{d}$ {Luminosity in the 0.5-10\,keV range, extrapolated from the best-fit model.} \\
  $^{e}$ {Emission and/or absorption-like features detected $<1$\,keV were modelled using a combination of Gaussian absorption lines. For justification and details see Section \ref{sec:xspec}. } \\
  $^{f}$ {Parameter frozen at total galactic H\,I column density \citep{2016A&A...594A.116H}.} \\
  $^{g}$ {There is considerably uncertainty in the distance estimation for this source. Radius is estimated for a distance of 7\,kpc \citep{2004ApJ...616L.139W}. Error bars are estimated from the 1\,${\sigma}$} errors for the best-fit value of $K_{\rm dBB}$. \\
  $^{h}$ {With exponential cutoff at high energies see Section \ref{sec:xspec} for details.} \\

 \label{tab:cont_fit}
\end{table*}

\label{sec:xspec} 
The X-ray spectral analysis was performed using the {\tt Xspec} spectral fitting package, version 12.9.0 \citep{1996ASPC..101...17A}. For the continuum fitting we used simple models consisting of the fewest possible free parameters. These were the \texttt{xspec} model \texttt{diskbb} for a multi-color disk black body (MCD), the spherical black body model (\texttt{bbodyrad}) for thermal emission components and a simple power-law model (with occasional exponential cutoff) for non-thermal emission. Despite their simplicity, all models are physically motivated. The MCD component is expected to originate in the accretion disk and the power law simulates non-thermal emission from a hot electron corona \citep[e.g.,][]{2007A&ARv..15....1D,2010LNP...794...17G}. The additional \texttt{bbodyrad} component was employed to model thermal emission from the NS surface, known as boundary layer emission \citep[e.g.][]{2001ApJ...547..355P}.

When detected, the Fe K${\alpha}$ emission line is modeled using a simple Gaussian model, based on which we estimate the EW. For consistency check, we estimated the EW of the emission line in two ways: a) For the broadband continuum fit. b) For a fit  of only the 4-8\,keV spectral range -- where the continuum is modelled using a smooth power-law. We found that both estimations were consistent, within 1${\sigma}$ error bars and we, therefore, opted to report the EW value of the broadband fits.
In case of non-detection we estimate the 1${\sigma}$ upper limit on the EW of a Gaussian emission line with a  width between 10-500\,eV and a centroid energy between 6.4-6.9\,keV. More specifically, we produce a range of EW upper limits of a grid of emission lines with the centroid energy in the 6.4-6.9\,keV with a step of 0.1\,keV and width within the 10-500\,eV range with a step of 20\,eV. We report the average value of all EW upper limit estimations. 

Due to the non-solar abundance of the accreted  material in UCXBs, their spectra are often interspersed with narrow absorption and/or emission-like features {in the soft ($<$1\,keV) X-rays \citep[e.g.,][]{2001ApJ...546..338P,2001ApJ...560L..59J}.} They are likely the result of emission by overabundant elements such as C and O but also pronounced absorption edges from  highly non-solar circum-source material. A meticulous study of these features would involve modeling the interstellar absorption by adding a secondary absorption component for the circum-source material assuming non-solar abundances of elements, along with high resolution analysis of the emission spectra (using e.g.~RGS). Such an endeavour is outside of the scope of this work. For this reason all absorption or emission-like residuals below ${\sim}2$\,keV where modelled, empirically using Gaussian absorption and emission lines and in some cases (see below) non-solar interstellar absorbers. No attempt was made to exact any physical interpretation of such features and they were only modelled in order to ensure a robust description of the observed spectra. In most  cases, dedicated analysis of these features has been done by previous authors. We refer to these works in all relevant segments of our text.

\subsection{Source specific comments}
\label{specific}

Below we briefly describe the spectral analysis of the sources in our UCXB catalog. When thermal emission is detected we convert the model normalization parameter into inner disk radius (for the \texttt{diskbb} model) and size of emitting region (for the \texttt{bbodyrad}). These estimations depend on the source distance -- and the error bar estimation is dominated by the distance estimation. All current distance estimations are presented in Table \ref{tab:sources_all}. The interstellar absorption was modeled using the improved version of the \texttt{ tbabs} code\footnote{http://pulsar.sternwarte.uni-erlangen.de/wilms/research/tbabs/} \citep{2000ApJ...542..914W}. The atomic cross-sections were adopted from \cite{1996ApJ...465..487V}.  We have divided the sources into two groups based on the presence or absence of the iron K${\alpha}$ emission line. To allow for a more concise reading experience, we do not describe the spectral fitting results of each individual source, but rather summarise the process for the whole sample. All best-fit values for the continuum fitting are presented in Table \ref{tab:cont_fit}. All estimations for the iron K${\alpha}$ emission line are tabulated in Table \ref{tab:LINES}.

In this table we use the notation {"Soft" for sources where thermal emission is either the sole component or has a 0.5-10\,keV flux at least three times higher than the non-thermal component, and "Hard" for spectra dominated by non-thermal emission (using the same rule). The two notations allude to the so-called "Soft" and "Hard states" of XRBs \citep[e.g. see review by][]{2010LNP...794...17G}. We note that the state transition of XRBs are much more nuanced than a simple division into soft and hard states (e.g., in our two "transitioning" sources the two components have comparable flux)} and that these characterisations are used to briefly denote the dominant spectral component in Table \ref{tab:LINES}. 

 \begin{figure*}
	\includegraphics[angle=-90, width=8cm]{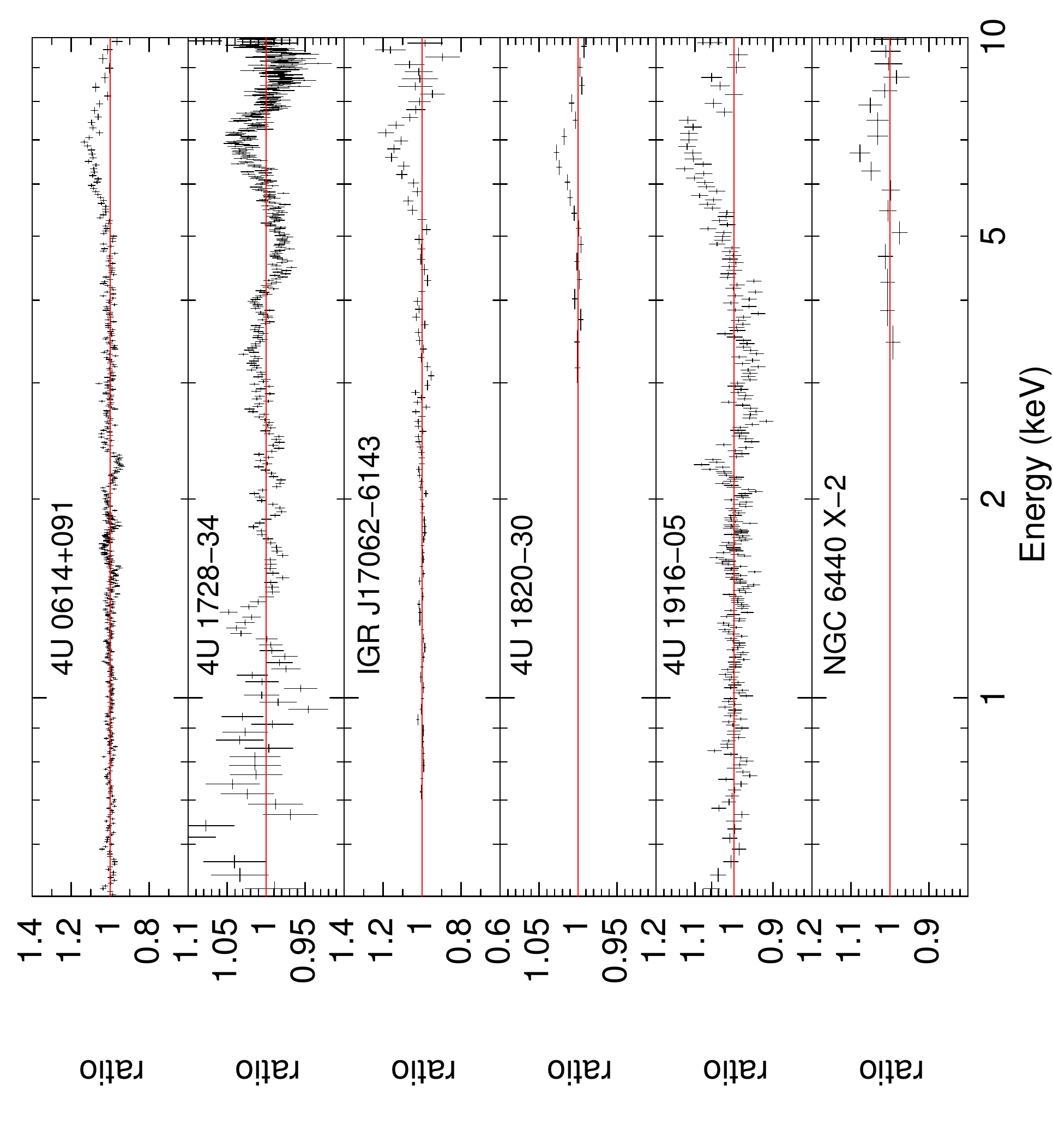}
    \includegraphics[angle=-90, width=8cm]{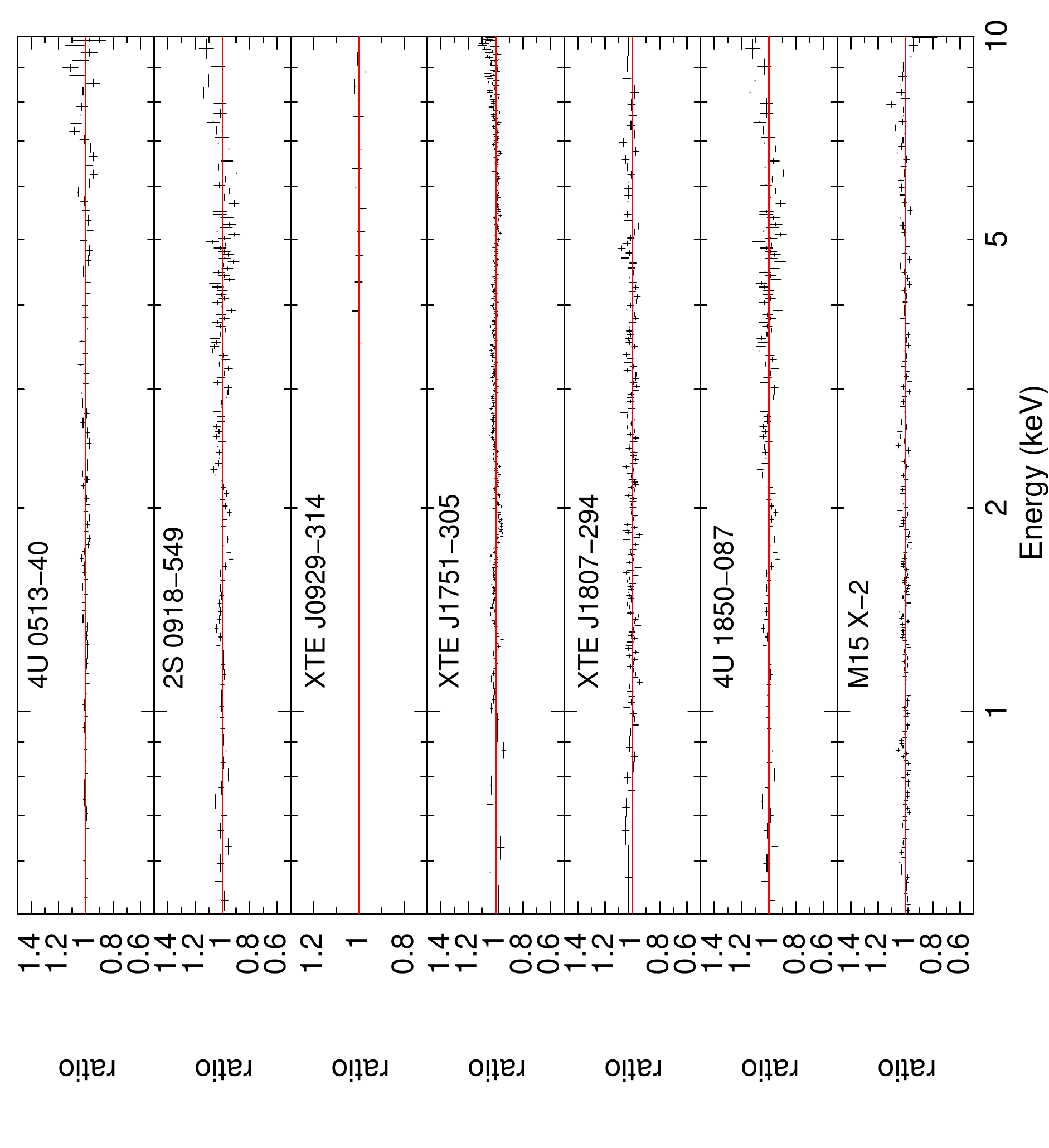}
    \caption{{\emph Left:} Data-to-model ratio plots for all sources with a detected Fe K${\alpha}$ emission line. The model includes all fit-required components (i.e. thermal and non-thermal continuum emission, soft  X-ray emission features), except the Gaussian component used to model the iron emission. 
    {\emph Right:}  Data-to-model ratio plots for all sources with no iron line detection. All sources are included except for 4U~1626-67. Due to its unique nature -- as an UCXB, X-ray pulsar -- the analysis of this source is presented separately. }
    \label{fig:LINES}
\end{figure*}
 
\begin{figure}
 \includegraphics[angle=-90, width=\columnwidth]{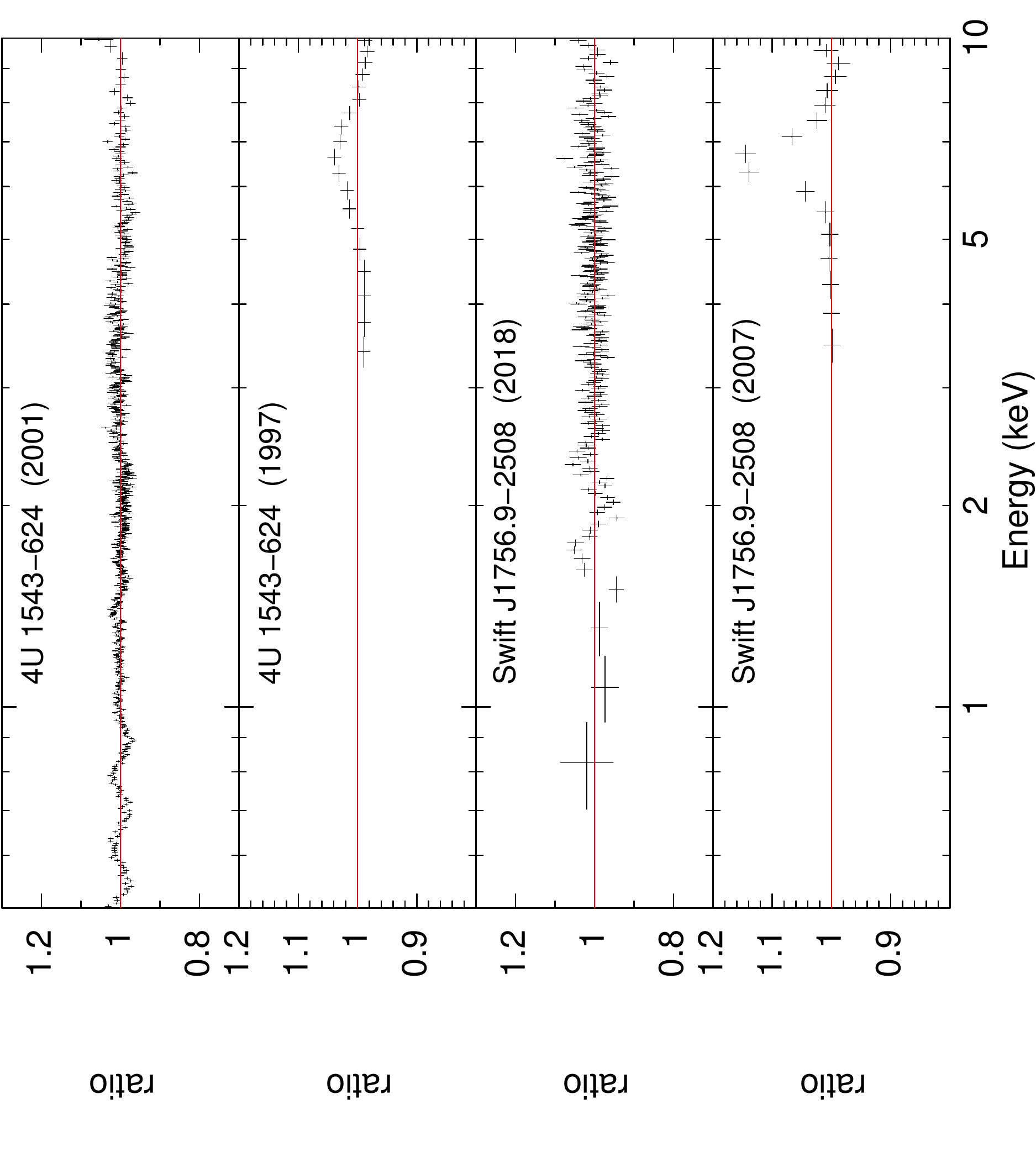}
 \caption{ Data-to-model ratio plots for the two sources with a variable Fe K${\alpha}$ emission line.}
 \label{fig:variable}
\end{figure}

\begin{figure}
 \includegraphics[angle=-90, width=\columnwidth]{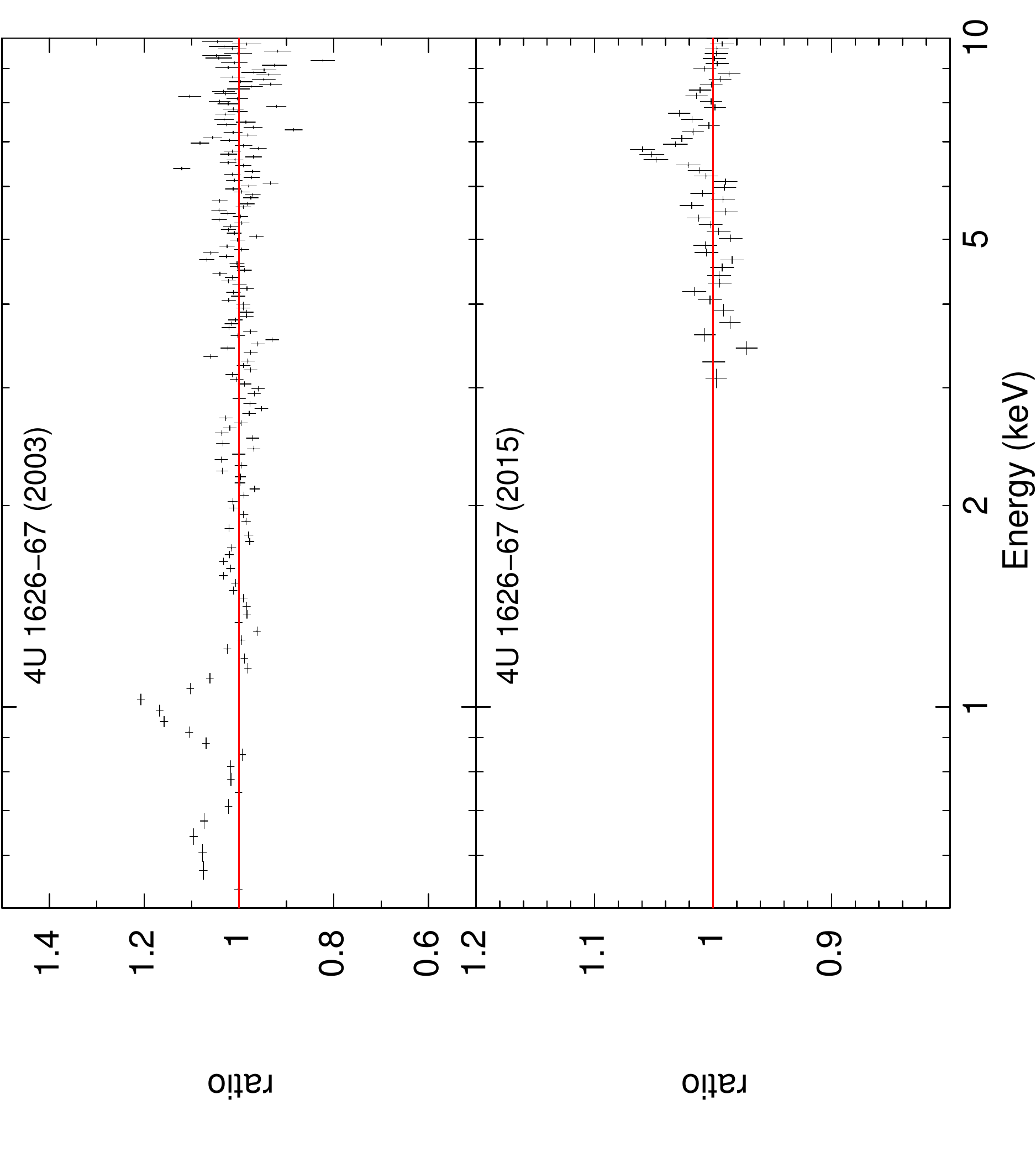}
 \caption{ Data-to-model ratio plots for the variable  Fe K${\alpha}$ emission line detection in X-ray pulsar 4U~1626-67.}
 \label{fig:1626}
\end{figure}

 \renewcommand{\headrulewidth}{0pt}
 \fancyhead[L,R]{}
\begin{table*}
 \begin{center}
\caption{Final results for the ultracompact X-ray binaries sample.}
\label{tab:LINES}
\begin{tabular}{|l|ccccccc|}
\hline
{Name of sources}   & Hard/Soft state& Iron line                   & Centroid energy      & Width                & EW\footnotemark[1]   & 1$\sigma$ upper limit\footnotemark[2]  & {Bursting activity\footnotemark[3]} \\
                    &                &                             & (keV)                & (keV)                & (eV)                 & (eV)           & \multicolumn{1}{c|}{}       \\ \hline
4U~0513-40          & Hard           & {\color[HTML]{FE0000} No}   & --                   & --                   & --                    & 5.59           & $35$ bursts ($1198$ obs)   \\
4U~0614+091         & Hard           & {\color[HTML]{34FF34} Yes}  & $6.59\pm0.09$        & $0.55\pm0.10$        & $63\pm6.19$         & --              & $17d\pm2$                   \\
2S~0918-549         & Soft           & {\color[HTML]{FE0000} No}   & --                   & --                   & --                   & 7.16           & $8$ bursts ($772$ obs)      \\
XTE~J0929-314       & Hard           & {\color[HTML]{FE0000} No}   & --                   & --                   & --                   & 19.7          & No bursts observed          \\
4U~1543-624         & Soft           & {\color[HTML]{FE0000} No?}  & --                   &  -                   & --                   & 2.56           & No bursts observed           \\
IGR~J17062-6143     & Soft/Transition& {\color[HTML]{34FF34} Yes}  & $6.71\pm0.11$        & $0.65\pm0.15$        &$237_{-104}^{+61.1}$ & --              &  $2$ bursts (5 observations)   \\
4U~1728-34          & Soft           & {\color[HTML]{34FF34} Yes}  & $6.58\pm0.03$        & $0.59\pm0.05$        &$76.1_{-8.20}^{+10.1}$& --              & $1067$ bursts ($4039$ obs)   \\
XTE~J1751-305       & Soft           & {\color[HTML]{FE0000} No}   & --                   & --                   & --                   & $ 5.87 $       & No bursts observed           \\
Swift~J1756.9-2508  &Soft            & {\color[HTML]{FE0000} No?}  & --                   & --                   & --                   & $4.95$          & No bursts observed           \\
XTE~J1807-294       & Soft           & {\color[HTML]{FE0000} No}   & --                   & --                   & --                   & $ 12.8 $       & No bursts observed            \\
4U~1820-30          & Hard           & {\color[HTML]{34FF34} {Yes}}& $6.52\pm0.07$        &$0.54_{-0.11}^{+0.16}$&$45.6_{-10.1}^{+16.5}$& --              & $\approx $2-4h                 \\
4U~1850-087         & Soft           & {\color[HTML]{FE0000} No}   & --                   & --                   & --                   & $ 6.38 $       & $4$ bursts ($1295$ obs)      \\
4U~1916-05          & Hard/Transition& {\color[HTML]{34FF34} {Yes}}& $6.65$               &$0.68_{-0.08}^{+0.09}$& $162{\pm}22.1$       & --              & $\approx$6.2h                 \\
4U~2129+11 (M15~X-2)& Hard           & {\color[HTML]{FE0000} No}   & --                   & --                   & --                   & $ 8.17 $       & $1$ burst ($162$ obs)         \\
NGC~6440~X-2        & Hard           & {\color[HTML]{34FF34} Yes}  & $6.71\pm0.20$        & $0.1$\footnotemark[4]& $75.1\pm20.6$        & --              & No bursts observed        \\
\hline
4U~1626-67          & ---             & {\color[HTML]{FE0000}No?}  & --                   & --                   & --                   & $ 14.2 $       & No bursts observed            \\
\hline

\end{tabular}
 \end{center}
\footnotemark[1]~{Computed from best fit value and  1$\sigma$ error bars of the line flux provided by the Gaussian model.}\\
\footnotemark[2]~{1$\sigma$ upper limit for the EW of the iron line are calculated by including a Gaussian with a fixed width of 0.1keV, centered between 6.4keV-6.9keV.}\\
\footnotemark[3]~{From \cite{2016ApJ...826..228J} and \cite{2003ApJ...595.1077C} (4U~1820-30) for recurrence times. From \cite{2013ApJ...767L..37D} and \cite{2017ApJ...836..111K,2018ApJ...858L..13S} for IGR~J17062-6143 and \cite{1990PASJ...42..633V} for 4U~2129+11. From \url{http://burst.sci.monash.edu/wiki/index.php?n=MINBAR.V09ReleaseNotes} for number of bursts, and from \url{https://personal.sron.nl/~jeanz/bursterlist.html} for the rest.}\\
\footnotemark[4]~{This parameter was frozen.}\\

\end{table*}

\subsubsection{Sources WITH a Fe K${\alpha}$ emission line}

In sources 4U\,0614+091, IGR\,J17062-6143, 4U\,1728-34, 4U\,1820-30, 4U\,1916-05 and NGC\,6440~X-2 we detect a prominent Fe K${\alpha}$ line with and EW ranging from ${\sim}$50 to ${\sim}$240\,eV.
In all sources the emission-like residuals on top of the data-to-model ratio plot, were prominent enough to allow for easy visual confirmation (Figure \ref{fig:LINES}, left panel). The addition of a Gaussian emission feature on all sources yielded a  ${\delta}{\chi}^2$ value of over 50 for 3 dof, indicating more than 6\,${\sigma}$ significance of detection. In several sources we also noted the presence of narrow residuals in the 1.7-2.5 kev range. They, most likely originate from incorrect modeling of the Si and Au absorption in the CCD detectors by the EPIC pn calibration \citep[e.g.][]{2010A&A...522A..96N}, which often results in the presence of emission and/or absorption features at ${\sim}$1.84\,keV, ${\sim}$2.28\,keV (M${\beta}$) and 2.4\,keV (M${\gamma}$). The features are not very pronounced and do not affect the quality of the fit or the aim of our analysis. Therefore, we have ignored them, but their respective energy channels are still included in our fits. 

In sources 4U\,0614+091 and 4U\,1728-34 we also detected  emission (at ${\sim}$0.5\,keV)  and absorption-like (at ${\sim}$0.7\,keV) features which where modelled using Gaussian lines.
In 4U\,1916-05 we further detect narrow absorption-like features at ${\sim}$6.65 and ${\sim}$6.95\,keV that are consistent with resonant absorption from Fe{\small XXV} and Fe{\small XXVI} ions, respectively, in agreement with the findings of  \cite{2006ApJ...646..493J}, using Chandra data. The best-fit values for the continuum are presented in Table \ref{tab:cont_fit} and for the iron K${\alpha}$ emission lines in Table \ref{tab:LINES}, along with the bursting activity (or lack thereof) of each source.

\subsubsection{Sources WITHOUT a Fe K${\alpha}$ emission line}

We found no evidence of iron line emission for sources 4U\,0513-40, 2S\,0918-549, XTE\,J0929-314, XTE\,J1751-305, XTE\,J1807-294, 4U\,1850-087, M15\,X-2 (4U\,2129+11\footnote{Not to be confused with neighboring source 4U\,2129+12}). For all sources we estimated the 1${\sigma}$ upper limits on the EW emission line, which we tabulate in Table \ref{tab:LINES}.
We also noted a prominent emission-like feature at ${\sim}$1\,keV in source XTE~J1807-294. Data-to-model ratio vs energy plots are presented in Fig.~\ref{fig:LINES}, right panel.

\subsubsection{Sources with a variable Fe K${\alpha}$ emission line}

In the 2001 \xmm observation of 4U~1543-624 no iron emission was detected in the 6-7\,keV region, with a tight EW upper limit of ${\sim}$3\,eV. The source was in the soft state, characterized by thermal emission which we model using a combination of MCD and spherical black body models.  Analysis of older (1997) RXTE observations, revealed the presence of a prominent  Fe K${\alpha}$ emission line \citep[see also analysis by][]{2003A&A...397..249S} for a similar continuum.  Swift~J1756.9-2508 was also dominated by thermal emission with no iron emission line detected with an 1${\sigma}$ upper limit of ${\sim}$5\,eV on the EW of a Gaussian emission line. However, in previous outbursts in 2007 and 2009 a very prominent iron emission line was detected in RXTE spectra of Swift~J1756.9-2508 \citep[see][for the iron emission in the 2009 observation]{2010MNRAS.403.1426P}. We have also analyzed the 1997 RXTE observation of 4U~1543-624 and the 2007 of Swift~J1756.9-2508, which we present in Table \ref{tab:appe} in the appendix. The variability of the line in the two sources is mentioned here and briefly discussed in Section~\ref{sec:discu}, but is thoroughly examined in a separate publication (Koliopanos et al.~accepted by MNRAS).

Apparent, complex residuals were detected in the sub-1\,keV spectral region of 4U~1543-624 and were modelled using two emission and one absorption Gaussian component. Due to  degeneracy between the interstellar absorption column density value and the parameters of the absorption lines, the column density was frozen to its galactic value (indeed, proper modelling of the line-of-sight absorption in 4U~1543-624 requires additional local absorption with extra-solar abundances, see e.g.~\citealt{2001ApJ...560L..59J}). Best-fit values of the spectral continuum of the sources are found in Table~\ref{tab:cont_fit} and EW upper limits of the Fe K${\alpha}$ line, along with bursting activity in Table \ref{tab:LINES}. The variable presence of the iron line of the two sources is depicted in their data-to-model ratio vs energy plots in Fig.~\ref{fig:variable}.
 
The third source with apparent iron line variability is 4U~1626-67 which is unique among UCXBs, as it is the only source powered by a highly magnetized (B${\sim}10^{12}$\,G) NS and is an X-ray pulsar. In this analysis we used a 2003 \xmm observation. The source continuum was fitted with a power-law with a spectral index of ${\sim}$0.8 and an exponential cutoff at ${\sim}$8\,keV and an MCD component with kT${\sim}$0.4\,keV. Narrow emission-like feature were also detected at  ${\sim}$0.6\,keV and  ${\sim}$1\,keV and were modelled using Gaussian emission lines. No iron line emission was detected during this observation. Nevertheless, after its 2010 luminosity increase and torque reversal, a faint emission Fe K${\alpha}$ emission line was detected in RXTE and Chandra data \citep[e.g.][]{2016MNRAS.456.3535K}. The \citeauthor{2016MNRAS.456.3535K} findings were confirmed by a 2015 \nus observation \citep[e.g.][]{2019ApJ...878..121I}. We also analyze this observation and tabulate our findings in Table \ref{tab:appe} in the Appendix. In the \nus observation (which was also taken during the more luminous spin-up era) we detect a moderately bright iron emission line with an EW value of ${\sim}30$\,eV.

We note that due to the nature of this source -- and despite our selection of the MCD model for the thermal emission in both observations -- it is likely that it does not originate in the accretion disk and the  iron line emission line is not the result of the standard coronal-disk reflection scenario modeled in \cite{2013MNRAS.432.1264K}. For these reasons, this source is discussed separately in Section~\ref{sec:discu} and its data-to-model ratio plot appears separately in Figure~\ref{fig:1626}, for both epochs. One of the most well monitored X-ray binaries, 4U~1626-67 has never been observed to produce X-ray bursts.

\section{Discussion}
\label{sec:discu}

\subsection{The spectral continuum}

We have revisited  X-ray observations of all known UCXBs. Our goal was not to retrace previous spectral analyses, but rather to model the spectral continuum in a consistent and physically motivated manner,  in order to draw conclusions on the presence or absence of the primary X-ray reflection feature in { X-ray spectra: the iron \ka line. We then use our findings  as a diagnostic} of the chemical composition of their disk and donor star. Our analysis of the spectral continuum confirms a physically credible interpretation of the spectral shape  for NS-XRBs, that has been noted in numerous previous studies \citep[e.g.][]{2001AdSpR..28..307B,2007ApJ...667.1073L}. From our sample of the sixteen known UCXBs we found eight sources in the "Soft" state, i.e. dominated by hot two thermal component (for sources 2S~0918-549 and IGR~J17062-6143 a faint power law tail is also detected), while the other eight were found in the "Hard" state, dominated by non-thermal emission and a cool MCD component (4U~1916-05 also features a hot thermal component, signifying a state transition).

During the soft state of accreting highly magnetized neutron stars, the accretion disk is expected to reach  the NS surface. Indeed, all sources in these state yield best fit values for the inner disk radius in the 10-15\,km range. A hot accretion disk extending to the surface of the NS. Furthermore, in all spectra we detect a secondary hotter thermal component (kT${\gtrsim}1$\,keV) consistent with emission from the hot, optically thick boundary layer expected to form on the surface of the NS during this state. Non-thermal tails during this state are usually detected in spectra from telescopes that extend at energies above 10\,keV. { The spectra of all sources found in the hard state were dominated by non-thermal emission, but also featured a less pronounced, but still significant soft MCD component.  The accretion disk temperature of hard-state sources is considerably lower than that of soft-state sources, and the inner radius ten to hundred of times larger.} These estimations are consistent with theoretical predictions of a truncated disk in hard state XRBs. The realistic quantities we derive from our best-fit values, increase the confidence in out analysis of the spectral continuum and consequently our estimations for the EW of the Fe \ka line. 

\subsection{Iron emission and X-ray diagnostics}

We have detected prominent and broad iron \ka emission lines in six of the known UCXBs (see Figure \ref{fig:LINES}, left). For all sources for which the emission line was detected -- and for which additional data with adequate statistics were available (see appendix for extra observations that were checked) -- we determined the stable presence of the emission line in their spectra. Another seven sources in the UCXB catalog showed no evidence of iron line emission (Figure \ref{fig:LINES}, right) with tight upper limits on iron emission, in most case EW$<$10\,eV. Similarly to the previous sample, all other available observations were inspected in order to confirm the absence of the Fe \ka line. The remaining three known UCXBs exhibited a variability in the presence of the iron line (Figures \ref{fig:variable} and \ref{fig:1626}). 

\subsubsection{Non detection}

With the help of our Monte Carlo simulation from \cite{2013MNRAS.432.1264K}, we can use the results of our analysis on the iron line detection, in order to evaluate the chemical composition of the accretion disk and donor star.  The code models fluorescence \ka and K${\beta}$ lines for elements between Z=3 and 30 and estimates their EWs with respect to the total emission, which is a combination of both the primary and the reflected component. For this work we have updated our code with the abundances by \citep[][]{2000ApJ...542..914W}\footnote{With abundances from \cite{1989GeCoA..53..197A} for elements not included in \citeauthor{2000ApJ...542..914W}} and atomic cross-sections from \cite{1996ApJ...465..487V}, for consistency with the \texttt{xspec} modelling. The code simulates the abundances of C/O-rich disk by "converting" H and  He into C and O. In practice we gradually reduce the solar-like abundances (by mass) of H and He, while increasing those of C and O by the same amount. The total number of nucleons is conserved, mass ratios of all other elements are kept fixed at their solar values and the C/O abundance ratio, also remains fixed. The position along this sequence of abundances can be give by the O/Fe ratio normalized to its solar value from \citep[][]{2000ApJ...542..914W}. The value of the O/Fe ratio corresponding to full "conversion" of H and He to C and O is 78. 

To evaluate the absence of the iron emission line in the spectra of our sources, we estimated the minimum value of the O/Fe ratio, in order to produce an Fe \ka line with an EW of 20\,eV, which is the highest upper limit in our source sample with no iron detection (source: XTE~J0929-314), for an incident power law with ${\Gamma}=1.9$ (best-fit parameter for XTE~J0929-314). Assuming an angle of 80$^{o}$ for the incident spectrum -- which yields the lowest iron line EW for a given O/Fe ratio -- we deduce that an O/Fe ratio of at least 12 is necessary in order to yield an iron emission line with EW$<$20\,eV. The O/Fe ratio increases to 25 for a face-on incident spectrum (0$^o$). Incident spectra with lower spectral index  or thermal primary emission (sources with thermal boundary layer emission at kT${\gtrsim}1$\,keV), require O/Fe${>}34$ in order for the iron line to be detected in their spectra at an EW$<$20\,eV. Based on these estimations, the disappearance of the iron emission line from the spectra of the seven UCXBs: 4U~0513-40, 2S~0918-549, XTE~J0929-314, XTE~J1751-305, XTE~J1807-294, 4U~1850-087 and 4U~2129+11 is the result of an O/Fe ratio at least an order of magnitude higher than the solar value, which can be interpreted as the result of a C/O or O/Ne/Mg donor in these sources.

\subsubsection{Persistent \ka emission line}

Six sources in the UCXB catalog feature a persistent iron emission line in their spectra, detected in multiple past observations (see Table \ref{tab:LINES} for the sources with Fe \ka line detection and Table A2 in Appendix for alternative observations that we have inspected. Based on the findings of our simulations, if the iron emission lines are the result of disk reflection -- further supported by the fact that all lines are considerably broadened -- the disk and donor star must be dominated by helium.  We also note two of the three sources that are also found in the soft-state feature an emission line with a considerably higher value of the EW (sources IGR~J17062-6143 and 4U~1916-05). Our estimations (see e.g.~Table 2 in \citealt{2013MNRAS.432.1264K}) indicate that soft-state sources produce emission lines with higher EW values (i.e.~in the 150-250\,eV range). This is primarily due to the shape of the incident spectrum -- i.e.~a power law with ${\Gamma}{\gtrsim}1.51$ vs a black body with kT${\sim}1.5-3$\,keV  and is  the result of higher available photons near the 7.1\,keV K-shell absorption edge of iron. However, harder power-law shaped incident spectra (e.g. ${\Gamma}<1$, usually found in accretion column emission) can produce Fe \ka line with similar EW to the black body incident spectrum. While, the seeming agreement between our observational analysis and the simulation results is encouraging, it must be considered very cautiously, as there are several other parameters involved, including the disk-corona geometry, viewing angle and disk ionisation state.   

\subsubsection{Variable presence of the Fe {\ka} emission line}

Examination of the rich observational archive of the UCXB source catalog revealed three sources for which the presence of the iron \ka line varies through the years. This is a very intriguing discovery, since the variability of this process -- i.e.~screening of the iron line by carbon and oxygen -- had been predicted by our initial analysis in \cite{2013MNRAS.432.1264K}. More specifically, in C/O dominated disks the screening efficacy of the two elements depends entirely on whether they are fully ionized or not. When oxygen (and therefore carbon) become fully ionized in the disk, the EW of the iron line is expected to return to its initial value, which is largely determined by the abundance of iron in the disk. In the \citeauthor{2013MNRAS.432.1264K} paper, simple analytical arguments were used to predict a luminosity dependence of the iron line variability. Under the assumption that the ionization state of the disk mostly depends on the accretion rate, it was predicted that a critical luminosity value for the variability of the iron line presence was at ${\sim}10^{37}$\,erg/s, { in the 0.5-10\,keV range}. Nevertheless,  sources 4U 1543-624 and Swift J1756.9-2508 were observed at similar luminosity  (${<}10^{37}$\,erg/s) and at similar spectral states withe the Fe \ka line being the only emission component that varied dramatically between observations. { Therefore, a detailed study of these two sources and the conditions of the iron line variability is needed in order to interpret the phenomenon. Such an analysis was indeed conducted and is presented in a separate work by Koliopanos et al.~(accepted by MNRAS). }

The third source with a variable presence of the iron line is 4U~1626-67. This is an exceptional source -- even within the already extraordinary UCXB population, as it is the only UCXB which has a highly magnetized NS accretor. It is a persistent X-ray pulsar with a known magnetic field of ${\sim}${a few}${\times}10^{12}$\,G, based on detection of cyclotron resonant scattering features, \cite[e.g.][]{2019ApJ...878..121I}. The iron line variability on this source was first reported by \cite{2016MNRAS.456.3535K}. Using archival data it was demonstrated that the emission line -- which is present during high-luminosity spin-up periods -- disappears when the luminosity drops, as the source goes into a spin-down epoch. It was further argued that this behavior was due to a geometric effect, the result of change in the emission diagram of the accretion column, which in turn is luminosity dependent  \citep[see e.g. Fig.1 of][and references therein]{2007A&A...472..353S}. 

In this work we revisited the \xmm observation for spectral analysis during the spin-down period and the recent \nus observation taken during the current spin-up phase. We note that our use of a MCD component for the modelling of the soft thermal emission does not necessarily signify emission from the inner part of an accretion disk heated due to viscous dissipation of gravitational energy. Indeed, the best fit value of the inner disk radius for the MCD model is of the order of 10-80\,km for the two epoch; more than two orders of magnitude smaller than the expected size of the NS magnetosphere for the strength of its magnetic field and the source luminosity (assuming 20\% efficiency; see also estimations in \citealt{2016MNRAS.456.3535K}). It is very likely that the thermal component originates in optically thick material in the vicinity of the magnetosphere, which is further heated due to illumination from the central source emission -- a phenomenon that is often detected in X-ray pulsars (see \citealt{2000PASJ...52..223E} and \citealt{2004ApJ...614..881H} for a detailed description of this configuration and \citealt{2018A&A...614A..23K} for a recent observational example).

Based on this scheme, the reasoning (that was also put forward in \citealt{2016MNRAS.456.3535K}) is that the disappearance of the iron line during the low-luminosity era, is due to  significantly reduced illumination of the optically thick magnetospheric material, which in turn is the result of the changing beam pattern. This is further highlighted by the less luminous and significantly cooler soft emission, during the low-luminosity \xmm observation (Table \ref{tab:sources_all}). The evolution to a fan-beam pattern emission, results in full illumination and heating of the optically thick material, and the appearance of the iron line, during the high luminosity phase, as observed with \nus. We ran our code for a beam with a 20${^o}$ opening angle aimed at an optically thick rectangular slab using realistic abundances for a C/O WD donor, predicted by \cite{2001A&A...375...87G}, and the atomic cross-sections of \cite{1996ApJ...465..487V}. If the material is illuminated in a face-on configuration we can reproduce an emission line with EW of ${\sim}$25\,eV, for an incident spectrum with ${\Gamma}{=}0.7$, for the C/O-rich material. Increasing the inclination angle between the incident beam and the slab -- thus reducing the reflected fraction -- results in the complete disappearance of the emission line. Based on this analysis, one can argue that 4U~1626-67 has a C/O-rich donor. Nevertheless, the premise of He-rich donor and iron line variability that is entirely due to the geometrical configuration of the source, cannot be ruled out.

\subsection{Type I X-ray bursts and optical spectroscopy}

We made an effort to comprehensively report the bursting activity of all sources in the UCXB catalog, presented in Table \ref{tab:LINES}. The recurrence rate of type I X-ray bursts can also indicate the chemical composition of the accreted material. Frequent bursts can be attributed to accretion of helium dominated material and/or residual hydrogen. On the other hand, sparse or no bursts -- particularly in persistent sources -- can be considered indicative of C/O rich material with only limited amount of residual helium and/or hydrogen. In addition to their regularity, the characteristics of thermonuclear X-ray bursts -- predominantly their duration and light-curve shape -- are heavily dependent on the chemical composition of the accumulated material which in turn affects the nuclear burning. For instance, hydrogen burns to helium via the CNO cycle and helium burns to carbon through the triple-alpha process.  In broad terms, type I X-ray bursts can be divided into three categories based on their duration. Short bursts (${\sim}$10-100\,s, associated with surface, unstable H and He burning, see reviews by \citealt{1993SSRv...62..223L} and \citealt{2006csxs.book..113S}), intermediate bursts (${\sim}$15-40\,m, most likely powered by the burning of a thick layer of helium, e.g.~\citealt{2006ApJ...646..429C}) and the so-called superbursts (${\sim}1$\,day, powered by carbon burning, e.g.~\citealt{2008A&A...479..177K}). 

All sources in the UCXB catalog -- for which no iron line is detected in their spectra (including the variable sources) -- exhibit sparse type I X-ray bursts or no bursts at all. The source  with most bursts detected is 4U~0513-40 (35 bursts detected in over a decade of monitoring). Indeed, 4U~0513-40 is  Out of the 10 sources with no iron emission features, six are persistent sources (4U~0513-40, 2S~0918-549, 4U~1543-624, 4U~1850-087, 4U~2129+11 and 4U~1626-67) and four are transient (XTE~J0929-314, XTE~J1751-305, Swift~J1756.9-2508, XTE~J1807-294). On the other hand, the five out of six UCXBs with prominent and persistent iron emission lines are frequent bursters with hundreds of type I X-ray bursts recorded and some with estimated recurrence rates in the order of days or hours. Four of these five sources, are persistent emitters (4U~0614+091, 4U~1728-34, 4U~1820-30, 4U~1916-05), while the fifth (IGR~J17062-6143) is a transient source which has produced type I X-ray bursts during both of its outbursts. The sixth UCXB with an iron line detection is NGC~6440~X-2, which is a transient source. No bursts have been observed from this source, but it is important to note that NGC~6440~X-2 exhibits faint outbursts with very short duration, that usually go undetected by all-sky monitors \citep{2010ApJ...712L..58A}, for this reason there are only few available observations for this source.

In addition to their high recurrence rates, the burst characteristics of sources 4U~1820-30  \citep[e.g.][]{1995ApJ...438..852B,2003ApJ...595.1077C, 2008ApJS..179..360G}, 4U~1916-05 \citep[e.g.][]{2008ApJS..179..360G}, 4U~1728-34 \citep[e.g.][]{2008ApJS..179..360G, 2010ApJ...718..947M}, 4U~0614+091 \citep[e.g.][]{2012ApJ...760..133L} and IGR~J17062-6143 \citep[e.g.][]{2017ApJ...836..111K} are also consistent with helium burning. Most UCXB companion stars and accretion disks are too faint for optical spectroscopy.  4U~1916-05 has optical confirmation of a He-rich donor \citep{2006MNRAS.370..255N}, while tentative optical evidence for a C/O companion have been put forward for 4U~1626-67, 2S~0918-549 \citep[orbital period and low mass-transfer rate are also consistent with low-entropy C/O white dwarf donor,][]{2013ApJ...768..184H} and 4U~1543-624 \citep[][]{2004MNRAS.348L...7N,2006A&A...450..725W} and FUV for 4U+2129+11 \citep[C and He lines][]{2005ApJ...634L.105D}. On the other hand, contrary to our designation as a He-rich source, \cite{2004MNRAS.348L...7N} and \cite{2006A&A...450..725W} have proposed a C/O-rich donor for 4U~0614+091, based on the presence of C and O emission lines on optical spectrum from the VLT. Our re-classification, however, not only handily explains its persistent bursting activity, but is also in agreement with indications for a He-rich donor based on arguments on disk stability and evolution \citep{2013ApJ...768..184H}. FUV  \citep{2002AJ....124.3348H} and optical  \citep{2006A&A...450..725W} spectroscopy of 4U~1626-67 supports the presence of C/O-rich donor, in agreement to our classification. However, \cite{2013ApJ...768..184H} argue that the orbital period and persistent accretion of the source can best be explained by He-star models rather than a C/O-rich donor. 

\subsection{X-ray spectroscopy in the bibliography}

Meticulous scrutiny of the X-ray spectral continuum of 4U~0513-40, 4U~0614+091, 2S~0918-549 and 4U~1850-087 -- using moderate (ASCA, BeppoSAX and \xmm EPIC) and high resolution instruments (\cxo and \xmm grating spectrometers) -- have revealed the presence of unusually strong absorption edges, indicating highly non-solar composition of the local interstellar medium, attributed to C/O or O/Ne/Mg-rich donors \citep[e.g.][]{2001ApJ...546..338P,2001ApJ...560L..59J,2003ApJ...599..498J}. However, analysis of later observations, yielded standard solar-like abundances, revealing a temporal variability of the effect and thus indicating that the apparent non-solar abundances could be the result of fluctuations in the ionization state of the surrounding material \citep[e.g.][]{2005ApJ...627..926J,2010ApJ...725.2417S}, rather than the result of non-solar composition. { Low-energy (${\lesssim}1$\,keV) broad emission-like features} have also been detected in sources 4U~0614+091 \citep{2010MNRAS.407L..11M}, 4U~1453-624 \citep{2011MNRAS.412L..11M} and 4U~1626-67 \cite{2007ApJ...660..605K} and have been attributed to emission from overabundant carbon, oxygen and/or neon in the accretion disk material \citep[e.g.][]{2014MNRAS.442.1157M}. 

An interesting point is raised by the modelling of 4U~0614+091 by \cite{2014MNRAS.442.1157M}. In this work, they use a version of the  {\small XILLVER} X-ray reflection code  \citep{2010ApJ...718..695G,2011ApJ...731..131G,2013ApJ...768..146G} modified in order to simulate reflection by C/O-rich material (the authors refer to this version of their code as {\small ${\rm XILLVER_{CO}}$}). They successfully model the soft emission-like features as well as the iron \ka emission line and argue that the soft features are due to a C/O-rich accretion disk. This finding seemingly contradicts the findings of  our own MC simulations which clearly indicate the screening of the iron line (primarily) by the overabundant oxygen. However, it is important to note that {\small ${\rm XILLVER_{CO}}$} does not use a realistic H and He-poor chemical composition, but rather simulates it by artificially increasing the abundance of all elements apart from H, He, C and O by ${\times}$10 times and then further increases the abundances of C and O producing a grid of abundance patterns in order to fit the spectra of 4U~0614+091. In the two observations where their fit constrains the C/O abundance, they report a value ranging between 130 to 160${\times}$ their solar value. However, based on their scheme for mimicking the C/O-rich disk, their actual O/Fe ratio is 13-16${\times}$ its solar value; at least five times lower than the O/Fe ratio of an actual H-poor, C/O-rich disk (i.e.~${\sim}$80, \citealt{2013MNRAS.432.1264K}; \citealt{2014MNRAS.442.2817K} and this work). Furthermore, the accretion disk simulated by {\small ${\rm XILLVER_{CO}}$} in \citeauthor{2014MNRAS.442.1157M} is still largely dominated by hydrogen and helium, rather than carbon and oxygen -- which significantly affects the ionization state of the disk. On the other hand, while our own simulation features a realistic disk composition, it does not self-consistently estimate the ionization state of the C/O-rich disk. A combination of the merits of our own simulations and the highly sophisticated model of \citeauthor{2010ApJ...718..695G} will be attempted in a future project. 

{ Lastly, we note that in this work we have not included the notable UCXB candidate 47 Tuc~X9, whose observed radio/X-ray luminosity ratio is consistent with a black hole accretor \citep{2015MNRAS.453.3918M} and ${\sim}28$\,min periodic modulation of its X-ray emission strongly indicates an orbital period consistent with UCXBs. Subsequent, extensive analysis of the rich archive of X-ray observations of  47 Tuc~X9 provided further evidence strongly favoring the scenario that this is the first black hole UCXB \citep{2017MNRAS.467.2199B}. Spectral and temporal analysis of the X-ray emission by \citeauthor{2017MNRAS.467.2199B} indicates a C/O-rich WD donor star. The detection of strong oxygen lines below 1\,keV and the absence of an iron emission line, is in agreement with our predictions for C/O-rich donors. The absence of any type I X-ray bursts from this source is also consistent with our predictions (but it is also consistent with the black hole accretor, regardless of the chemical composition of the donor material). However, the low source luminosity ($L_{X}<10^{34}$\,erg/s) suggests a very low-accretion state that points to the presence of an optically thin inner accretion flow (with possible optically thin plasma emission) and an accretion disk that is truncated at a distance that is at least an order of magnitude larger than in our Monte Carlo simulation. Therefore, we would need to make significant modifications to our code, in order to make robust claims about the nature of 47 Tuc~X9. This is beyond the scope of this paper and is considered for a separate publication.}
 
\section{Conclusions}
\label{sec:conc}

The analysis of the X-ray spectra of all known UCXBs -- using the predictions of the X-ray reflection code developed in \cite{2013MNRAS.432.1264K} -- has enabled us to draw potential conclusions for the chemical composition of their accretion disks and donor star. We found that all UCXBs featuring a prominent iron \ka emission line in their X-ray spectra, that is persistent over multiple observations, also exhibit a rich and recurrent bursting activity. On the other hand, all UCXBs for which the iron line is not detected in all or most of their available spectra, also have few or no type I X-ray bursts detected, in over more than a decade of monitoring. Based on the strong correlation between the strength of the Fe \ka emission line and the abundance of carbon and oxygen in the accretion disk material -- demonstrated in \cite{2013MNRAS.432.1264K} -- and the correlation between the abundance of hydrogen and helium in the recurrence rate of short and intermediate type I X-ray bursts, we argue that there is a strong likelihood that all known UCXBs with persistent iron emission have He-rich donors, while those that do not, likely have C/O or O/Ne/Mg-rich donors. 

If confirmed, this result can have profound implications in our understanding of compact binary evolution especially since standard population synthesis models tend to produce an overwhelming majority of He-rich systems rather than C/O-rich systems \citep[e.g.][]{2010MNRAS.401.1347N}. Furthermore, recent hydrodynamic simulations of mass transfer in binaries indicate that NS's with CO or O/Ne WD donors lead to rapid NS-WD mergers and thus UCXBs with such donors would be very short-lived systems \citep{2017MNRAS.467.3556B}. Comparison of our conclusions with assessments based on optical and X-ray spectroscopy returned positive results, but also some tensions. We also found indications of the iron line variability in C/O-rich disks, predicted in \cite{2013MNRAS.432.1264K}. Overall,it becomes evident that an increase in the  sophistication of our X-ray reflection models is required concerning the problem of non-solar abundances and disk ionization in order to definitively confirm our current conclusions. We therefore only regard the results of this work as encouraging and intriguing and conclude that they certainly justify further exploration. 

\section*{Acknowledgements}
F.K.~is deeply indebted to Marat Gilfanov, whose guidance made this series of projects possible. F.K.~is also grateful for the collaboration of Lars Bildsten and Maria Diaz Trigo, in the works that preceded and led to this final project. Lastly, F.K.~extends his utmost gratitude to the referee and editor of this paper, for their contribution to its final form, but more importantly their patience during a prolonged pause in the refereeing process, caused by F.K., due to an unforeseen personal issue.

\section*{Data availability}

The observational data underlying this article are publicly available at the NASA HEASARC archive\footnote{https://heasarc.gsfc.nasa.gov/docs/archive.html}. Any details with regard to the Monte Carlo source code can be shared on reasonable request to the corresponding author.




\bibliographystyle{mnras}
\bibliography{general}




\appendix
\label{appe}
\section{Some extra material}

\begin{table*}
 \caption{Best fit parameters for the continuum emission spectra of the three UCXBs with variable iron line emission. 4U~1626-67 -- a high-B NS, X-ray pulsar -- is tabulated separately. All errors are in the 1$\sigma$ confidence range.}
 \begin{center}
\scalebox{0.8}{   \begin{tabular}{lcccccccccc}
     \hline\hline\noalign{\smallskip}
     \multicolumn{1}{c}{Source} &
          \multicolumn{1}{c}{nH} &
     \multicolumn{1}{c}{k${\rm T_{dBB}}$} &
     \multicolumn{1}{c}{${\rm R_{dBB}}^{a}$} &
     \multicolumn{1}{c}{${\rm \Gamma}$} &
     \multicolumn{1}{c}{${\rm K_{po}}^{b}$} &
     \multicolumn{1}{c}{k${\rm T_{BB}}$}&
     \multicolumn{1}{c}{${\rm R_{BB}}^{c}$} &
     \multicolumn{1}{c}{${\rm {L}^{d}}$} &
          \multicolumn{1}{r}{red. ${\chi^2/dof}$ }\\

      \multicolumn{1}{c}{} &
      \multicolumn{1}{c}{${\times}10^{22}$} &
      \multicolumn{1}{c}{keV} &
      \multicolumn{1}{c}{km} &
      \multicolumn{1}{c}{} &
      \multicolumn{1}{c}{[$10^{-3}$]} & 
      \multicolumn{1}{c}{} &
      \multicolumn{1}{c}{km} &
      \multicolumn{1}{c}{{$10^{36}{\times}${erg\,s$^{-1}$}}} &
      \multicolumn{1}{c}{} \\
      \noalign{\smallskip}\hline\noalign{\smallskip}

4U~1543-624$^{h}$ (1997)   & 0.29$^{f}$      & 0.59$_{-0.22}^{+0.31}$&${11.5_{-1.56}^{+7.06}}^{g}$ & 4.32$_{-1.22}^{+1.31}$ & 0.35${\pm}0.05$ & 1.53${\pm}0.01$        & 0.32${\pm}0.06$ & 6.64   & 0.64/38 \\
Swift~J1756.9-2508 (2007)  & 0.01$^{f}$      & 1.85$_{-0.28}^{+0.26}$&${6.2_{-1.35}^{+3.76}}$      & --                     &  --             & 4.15$_{-1.01}^{+0.82}$ & 0.10${\pm}0.05$ &  4.53  & 0.62/32 \\
 \noalign{\smallskip}\hline\noalign{\smallskip}
4U~1626-67     (2015)    & 1.00$^{f}$     &0.68$_{-0.22}^{+0.31}$    &9.72$_{-1.22}^{+1.31}$ & 1.02$\pm0.004$     & 30.6$_{-2.8}^{+3.2}$ & --                      & --                     &  3.04   & 1.05/838 \\
     \hline\hline\noalign{\smallskip}
     \end{tabular}   }
 \end{center}
  
  $^{a}$ {$R_{\rm dBB}$ (the inner radius of the accretion disk in km) is inferred from  {\texttt {diskbb}} model, by solving K$_{\rm dBB}$=${\rm(R_{\rm dBB}/{D_{10}})^{2}}\,\cos{i}$, for $R_{\rm dBB}$. `K$_{\rm BB}$' is the normalisation of the  \texttt{diskbb} model, ${\rm D_{10}}$ is distance in units of 10\,kpc and $i$ is the inclination. In all sources this was assumed to be 60$^{o}$, with the exceptions of 4U~1916-05 and  4U~0513-40 (where $i$ is set to 85$^{o}$), which are known edge-on viewed systems \citep[][]{2004A&A...418.1061B,2011MNRAS.414L..41F}. With the exception of XTE J19029-314 (for which the K estimation has considerably high $1\,{\sigma}$ error bars), errors in radius estimation are dominated by the 1\,${\sigma}$ errors in the distance estimation (see Table \ref{tab:sources_all}). }\\
  $^{b}$ {Power-law component normalisation constant: photons/keV/cm$^2$/s at 1\,keV} \\
  $^{c}$ {Size of the spherical black body component (\texttt{bbodyrad} model), estimated from K$_{\rm BB}$=${\rm {R_{\rm BB}}^{2}/{D_{10}}^2}$, where K$_{\rm BB}$ is the normalization parameter of the \texttt{bbodyrad} model.}\\
  $^{d}$ {Luminosity in the 0.5-60\,keV range, extrapolated from the best-fit model.} \\
  $^{e}$ {Emission and/or absorption-like features detected $<1$\,keV were modelled using a combination of Gaussian absorption lines. For justification and details see Section \ref{sec:xspec}. } \\
  $^{f}$ {Parameter frozen at total galactic H\,I column density \citep{2016A&A...594A.116H}.} \\
  $^{g}$ {There is considerably uncertainty in the distance estimation for these two sources. Radius is estimated for a distance of 7\,kpc \citep{2004ApJ...616L.139W} for 4U~1543-624 and 8\,kpc for 4U~1626-67 \citep[4-13\,kpc;][]{1998Natur.394..346C}. Error bars are estimated from the 1\,${\sigma}$} errors for the best-fit value of $K_{\rm dBB}$. \\
  $^{h}$ {With an additional "constrained" broken power law ($p_{1}{\leq}2.5$ and $E_{b}{=}20$\,keV). This empirical model for high energy tails of NS XRBs in soft state, is often used in broadband X-ray spectroscopy (see \citealt{2007ApJ...667.1073L} and Koliopanos et al. submitted for more details). In the table we tabulate the value of $p_{2}$ for the spectral index value.} \\

 \label{tab:appe}
\end{table*}

\begin{table*}
 \centering
 \begin{minipage}{108mm}
\caption[Best fit parameters]{Best fit parameters for iron line emission in the three UCXBs with variable line detection. 4U~1626-67 -- a high-B NS, X-ray pulsar -- is tabulated separately. The errors are 1$\rm \sigma$.}
\label{tab:cont2}
  \begin{tabular}  {lcccccc}
  \hline
Model parameter                           & 4U 1543-624 (1997)    & Swift~J1756.9-2508 (2007)    & 4U~1626-67 (2010)          \\
  \hline                                                                                                                     
{\em Iron Line}                                                                                                                     \\
  \hline                                                                                                                     
     Centroid E (keV)                     & 6.58$_{-0.33}^{+0.21}$    & $6.58\pm0.06$           & 6.76$\pm0.03$                \\
     Width $\sigma$(keV)                   &0.68$_{-0.36}^{+0.55}$    & $0.1^{b}$               &0.19$\pm0.04$               \\
     Flux$^{\it a}$                       & 95.1$_{-23.5}^{+34.5}$     & 42.6$\pm4.60$         &    13.4$_{-1.61}^{+1.69}$                  \\
     EW (eV)  			                  & 118$_{-24.3}^{+36.6}$     & $173\pm18.5$            &   30.1$\pm3.85$             \\
\hline                                                                                      
\end{tabular}
 \medskip

{  $^{\it a}$  ${\rm 10^{-5}\,ph\,cm^{-2}s^{-1}}$. \\
$^{\it b}$  Parameter frozen at minimum spectral resolution value. \\}

\end{minipage}
\end{table*}

 \begin{figure}
       \resizebox{\hsize}{!}{\includegraphics[angle=-90,clip,trim=0 0 0 0,width=0.8\textwidth]{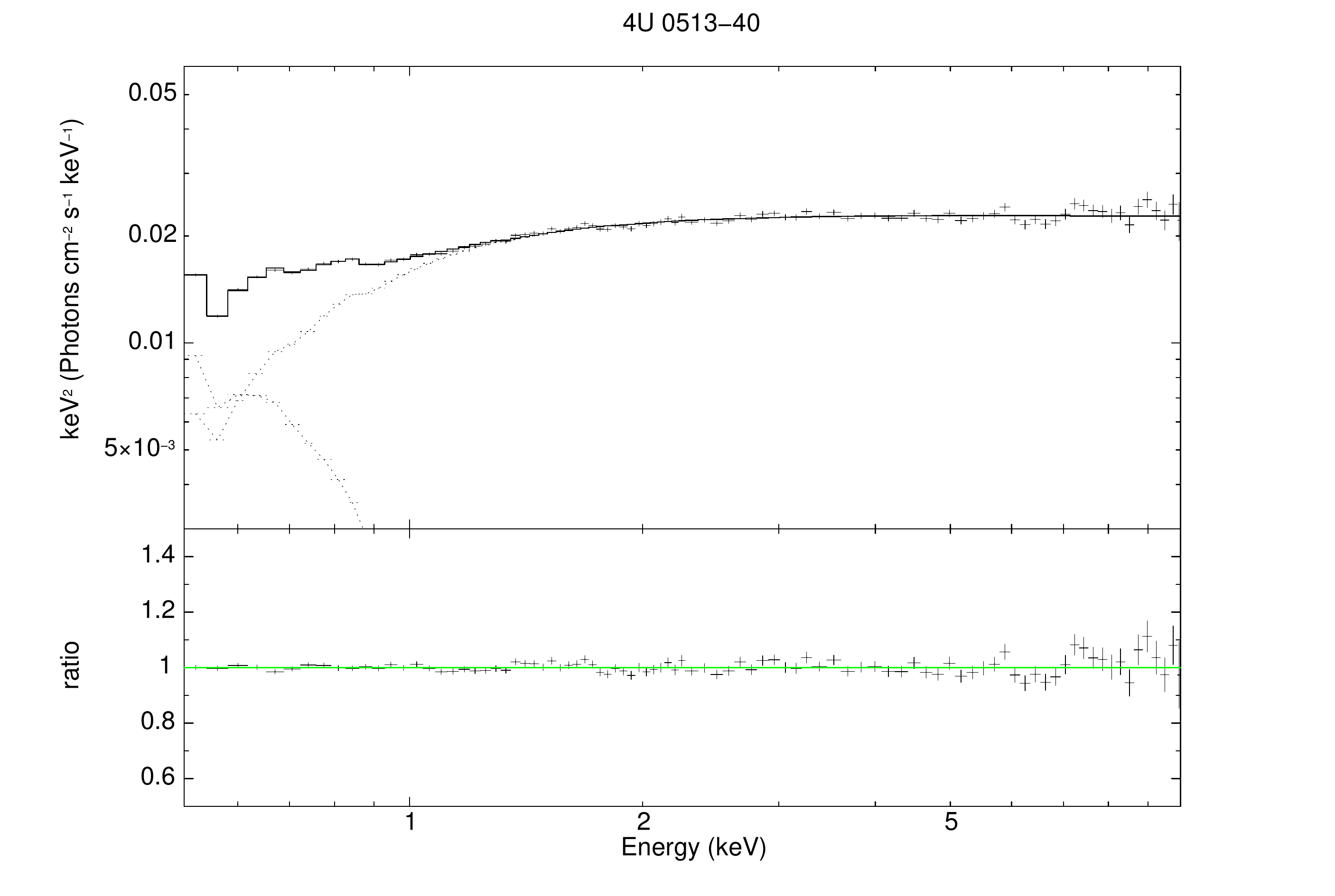}}
            \caption{4U~0513-40: Unfolded spectrum. Energy and data-vs-model ratio plot, for only the continuum model. }
   \label{fig:0513}
 \end{figure}
 
  \begin{figure}
       \resizebox{\hsize}{!}{\includegraphics[angle=-90,clip,trim=0 0 0 0,width=0.8\textwidth]{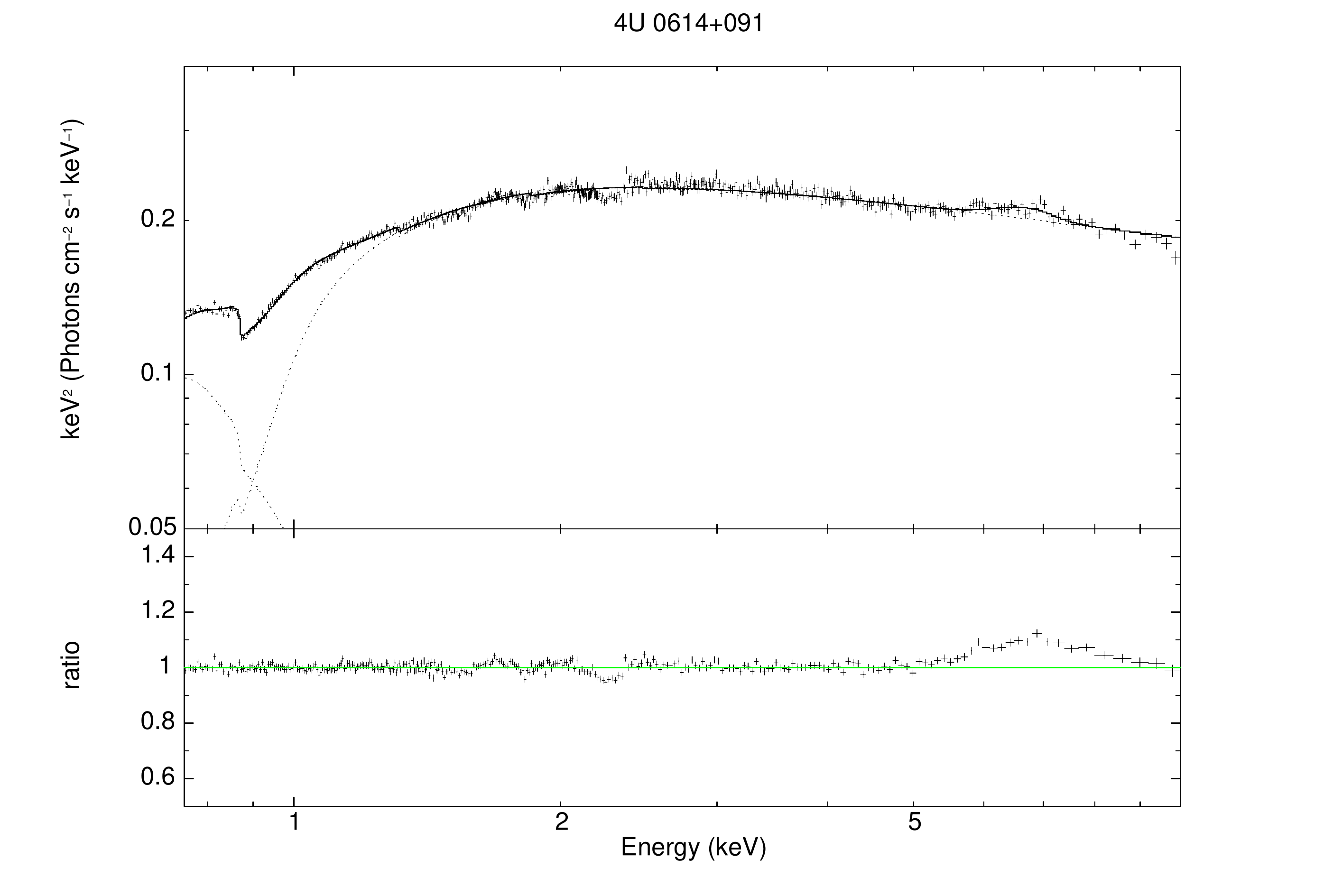}}
            \caption{4U~0614+091: Unfolded spectrum. Energy and data-vs-model ratio plot, for only the continuum model. }
   \label{fig:0614}
 \end{figure}
 
  \begin{figure}
       \resizebox{\hsize}{!}{\includegraphics[angle=-90,clip,trim=0 0 0 0,width=0.8\textwidth]{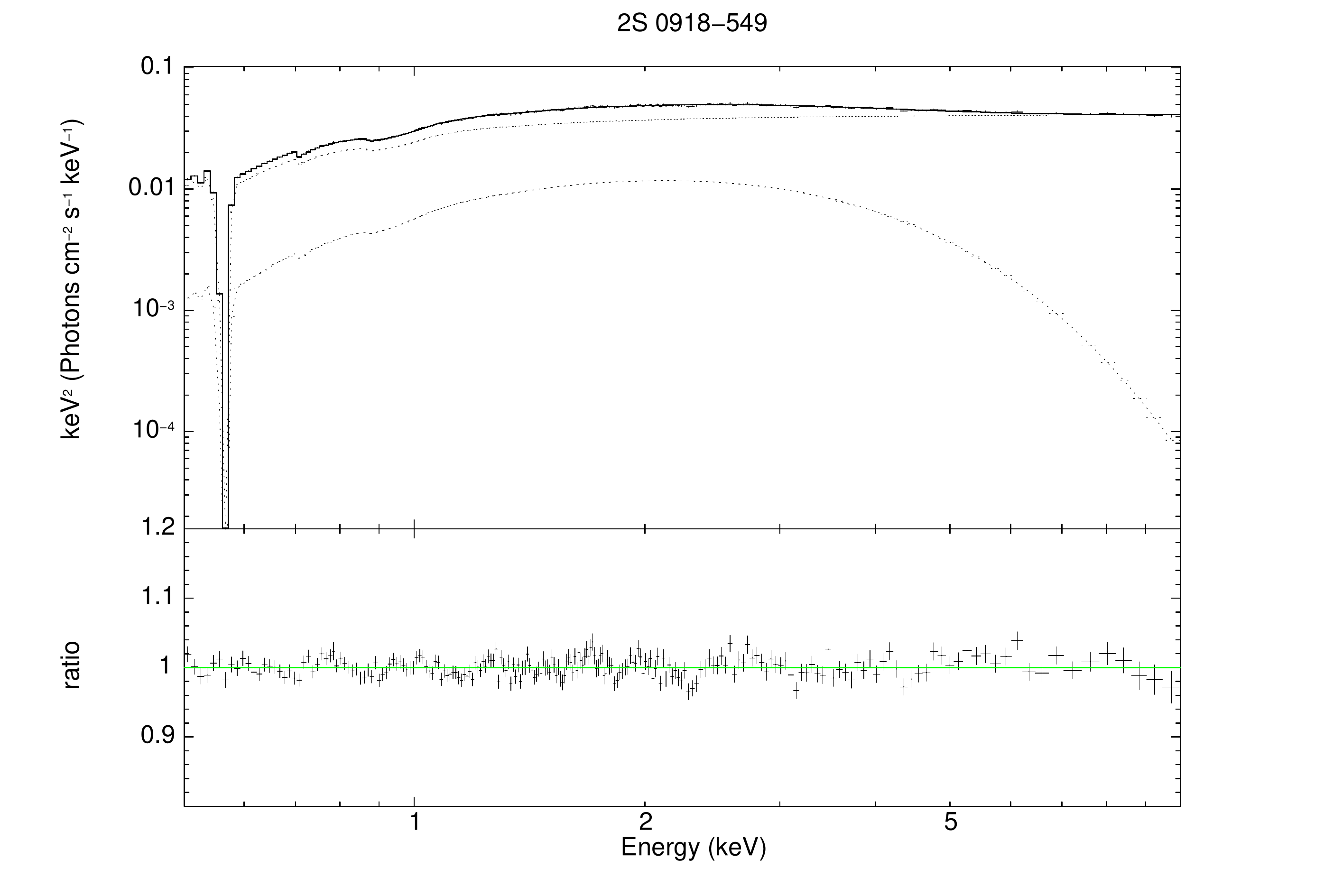}}
            \caption{2S~0918-549: Unfolded spectrum. Energy and data-vs-model ratio plot, for only the continuum model. }
   \label{fig:2S}
 \end{figure}
 
  \begin{figure}
       \resizebox{\hsize}{!}{\includegraphics[angle=-90,clip,trim=0 0 0 0,width=0.8\textwidth]{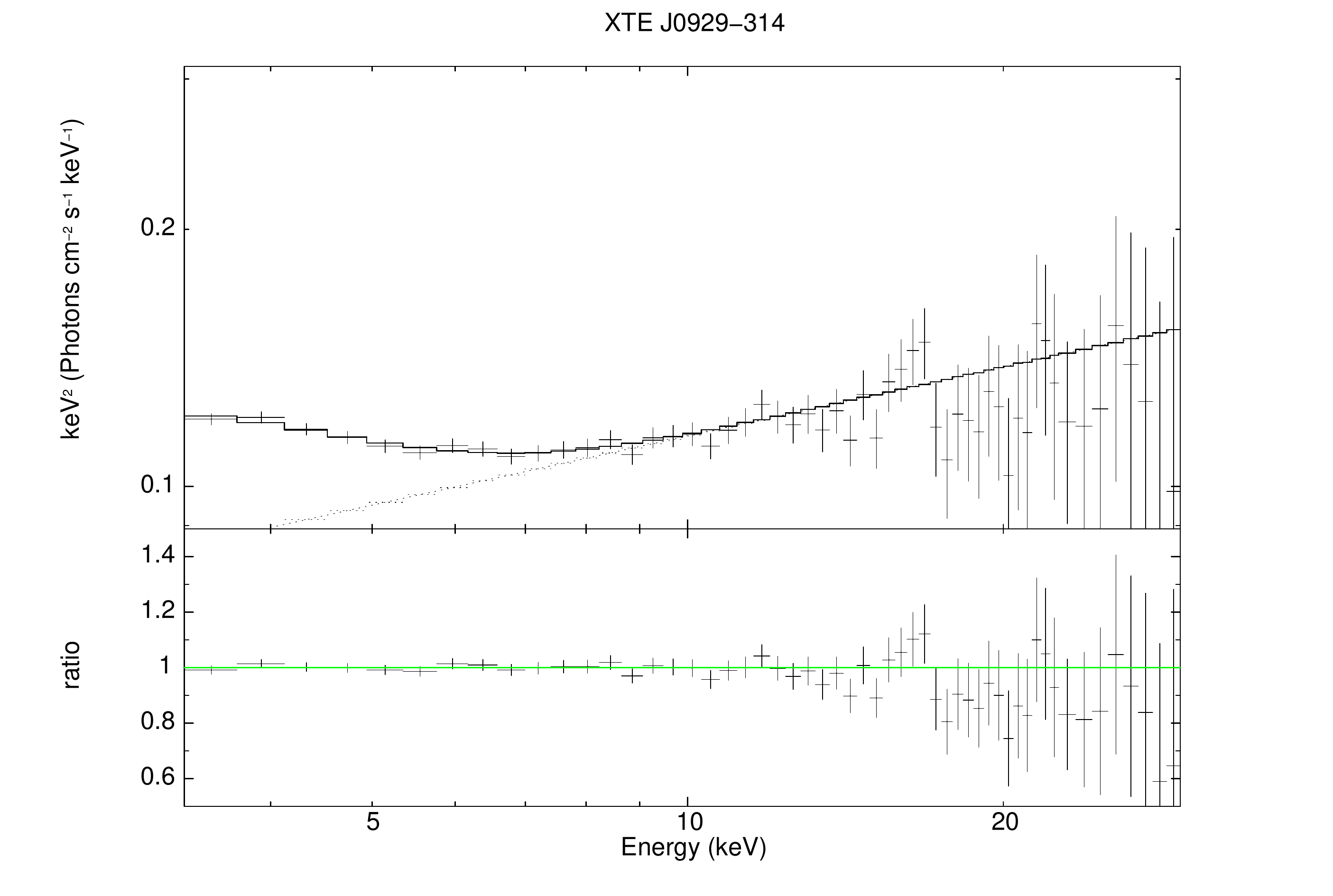}}
            \caption{XTE~J0929-314: Unfolded spectrum. Energy and data-vs-model ratio plot, for only the continuum model. }
   \label{fig:0929}
 \end{figure}
 
  \begin{figure}
       \resizebox{\hsize}{!}{\includegraphics[angle=-90,clip,trim=0 0 0 0,width=0.8\textwidth]{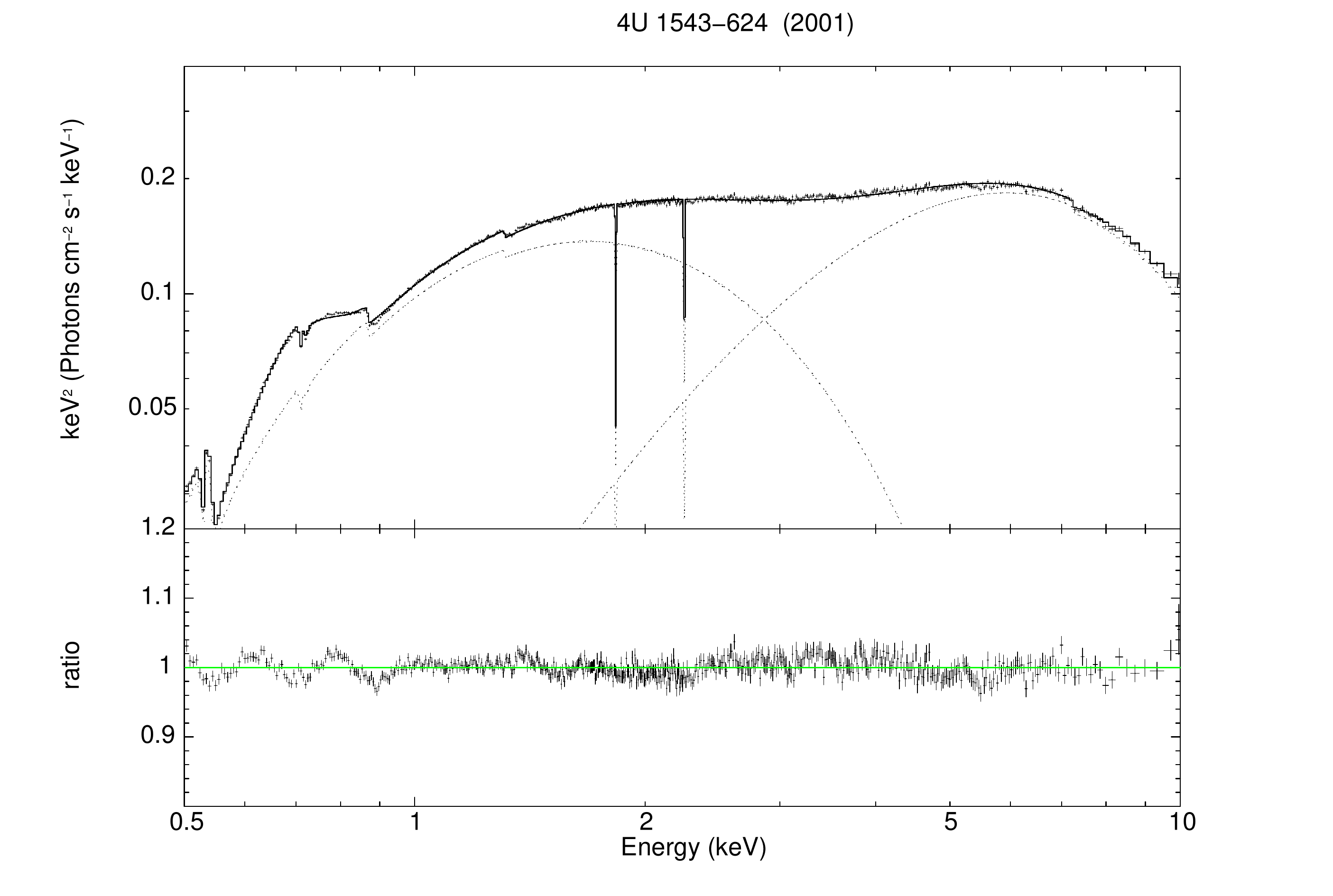}}
            \caption{4U~1543-624 (2001): Unfolded spectrum. Energy and data-vs-model ratio plot, for only the continuum model. }
   \label{fig:1543}
 \end{figure}
 
   \begin{figure}
       \resizebox{\hsize}{!}{\includegraphics[angle=-90,clip,trim=0 0 0 0,width=0.8\textwidth]{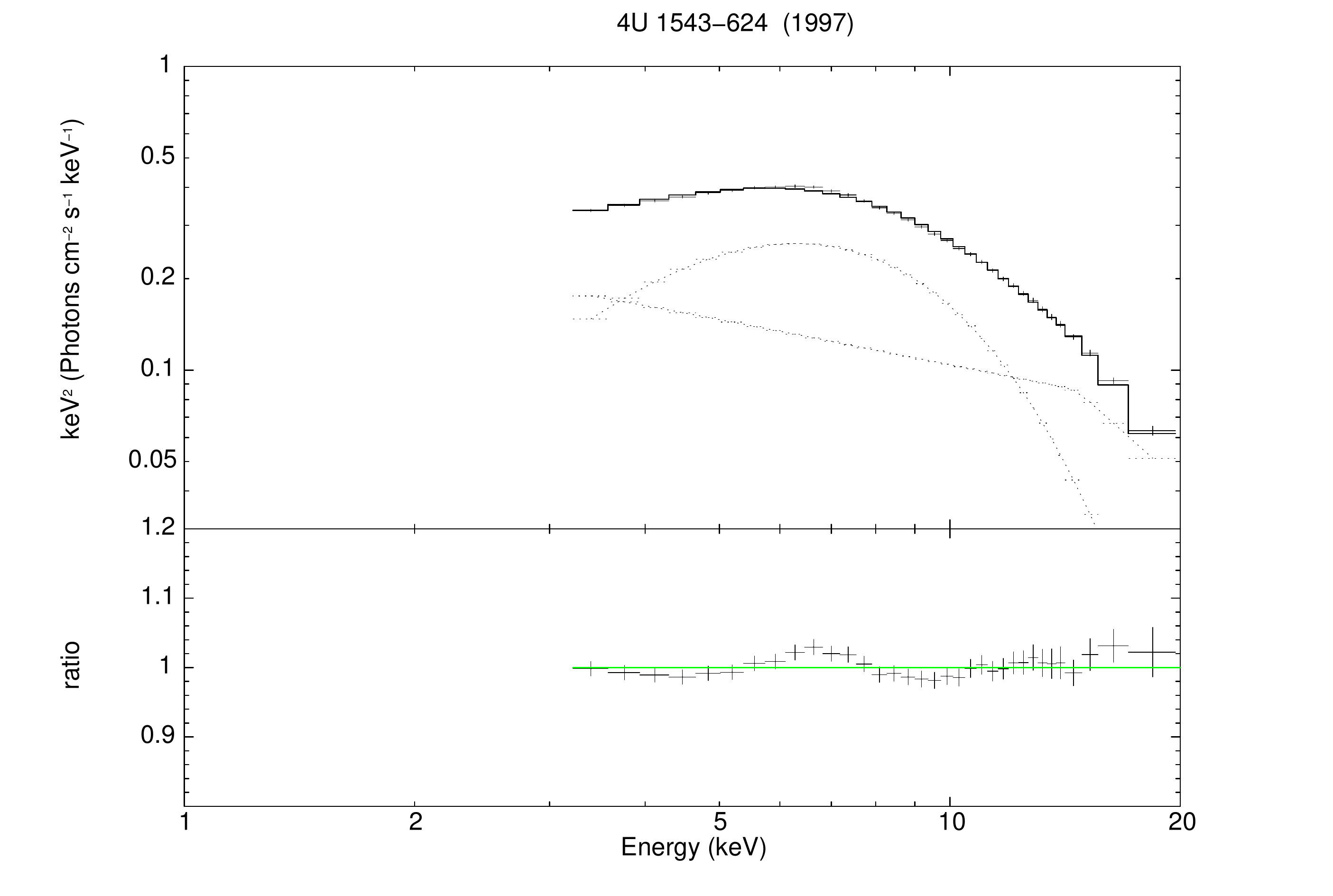}}
            \caption{4U~1543-624 (1997): Unfolded spectrum. Energy and data-vs-model ratio plot, for only the continuum model. }
   \label{fig:1543_RXTE}
 \end{figure}
 
  \begin{figure}
       \resizebox{\hsize}{!}{\includegraphics[angle=-90,clip,trim=0 0 0 0,width=0.8\textwidth]{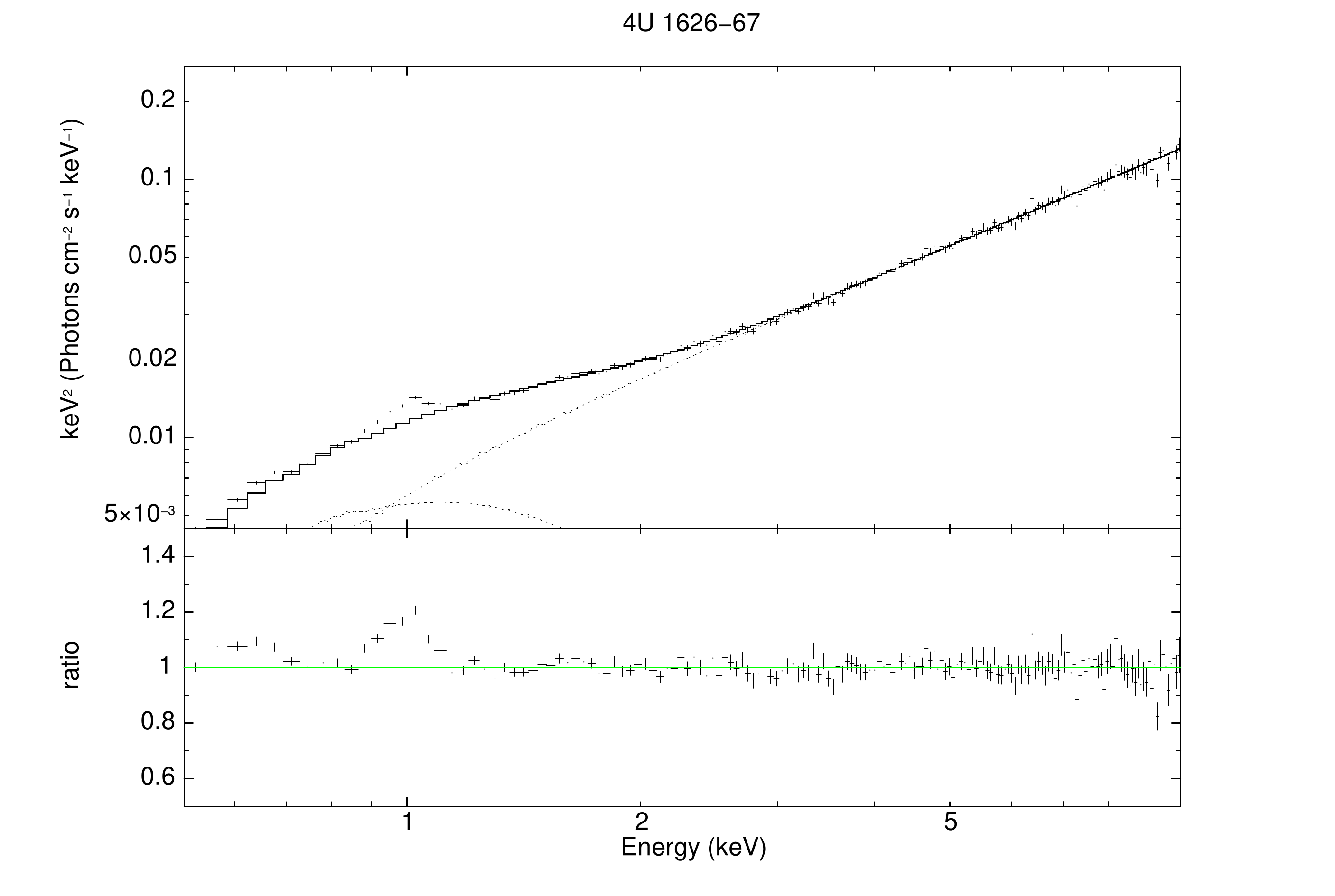}}
            \caption{4U~1626-67: Unfolded spectrum. Energy and data-vs-model ratio plot, for only the continuum model. }
   \label{fig:1626eeufs}
 \end{figure}

   \begin{figure}
       \resizebox{\hsize}{!}{\includegraphics[angle=-90,clip,trim=0 0 0 0,width=0.8\textwidth]{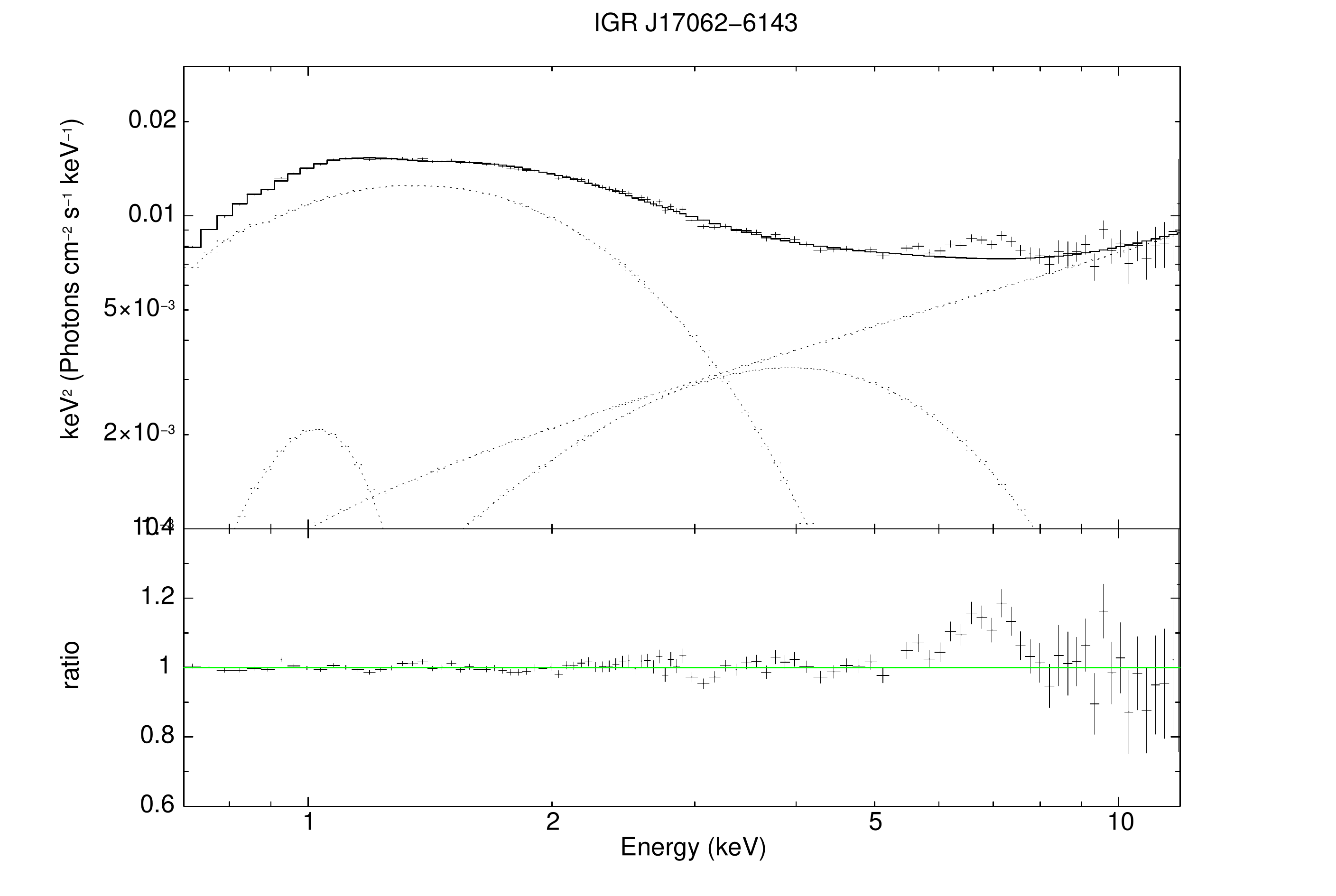}}
            \caption{IGR~J17062-6143: Unfolded spectrum. Energy and data-vs-model ratio plot, for only the continuum model. }
   \label{fig:IGR}
   
 \end{figure}
 
  \begin{figure}
       \resizebox{\hsize}{!}{\includegraphics[angle=-90,clip,trim=0 0 0 0,width=0.8\textwidth]{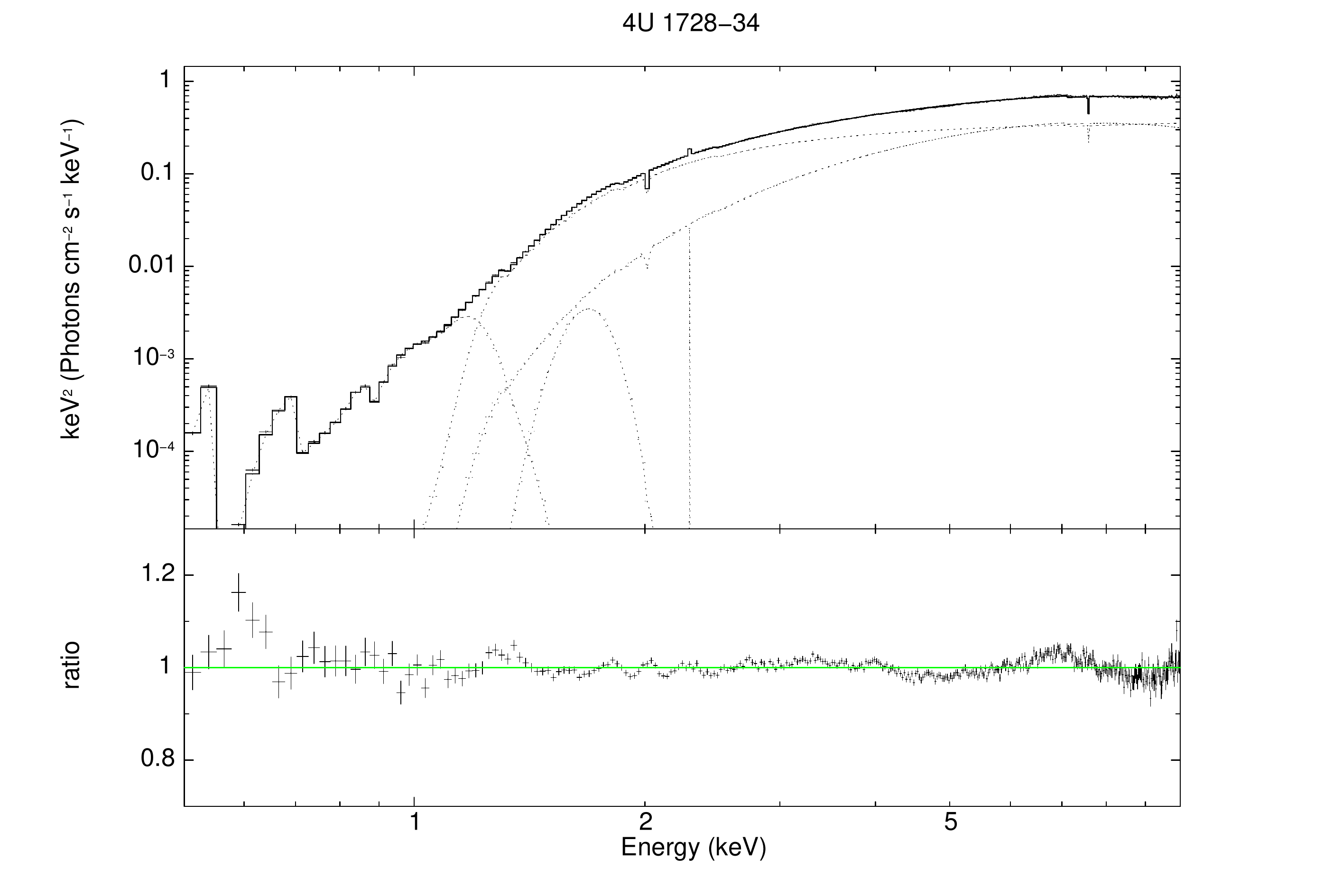}}
            \caption{4U~1728-34: Unfolded spectrum. Energy and data-vs-model ratio plot, for only the continuum model. }
   \label{fig:1728}
 \end{figure}
 
  \begin{figure}
       \resizebox{\hsize}{!}{\includegraphics[angle=-90,clip,trim=0 0 0 0,width=0.8\textwidth]{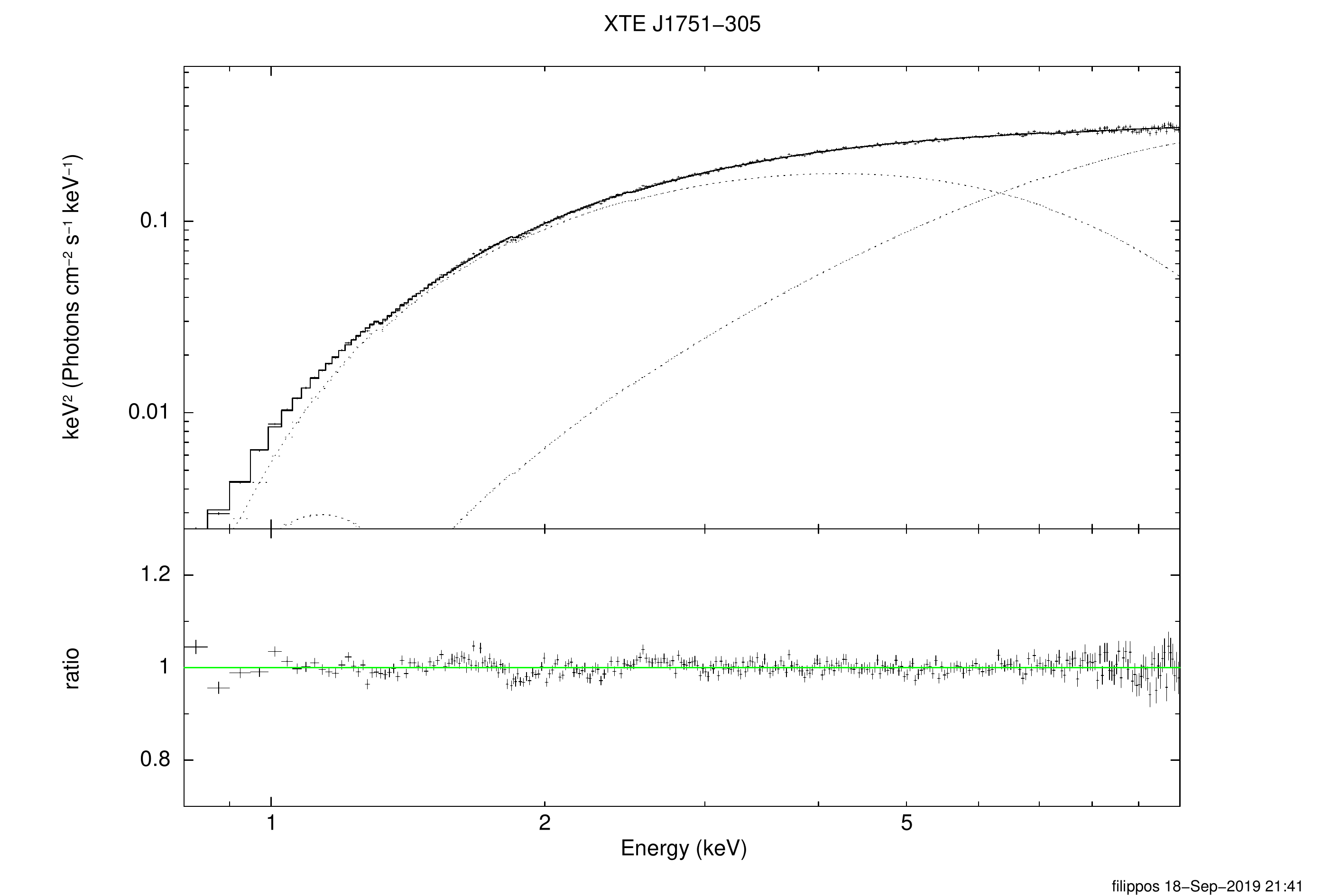}}
            \caption{XTE~J1751-305: Unfolded spectrum. Energy and data-vs-model ratio plot, for only the continuum model. }
   \label{fig:J1752}
 \end{figure}
 
  \begin{figure}
       \resizebox{\hsize}{!}{\includegraphics[angle=-90,clip,trim=0 0 0 0,width=0.8\textwidth]{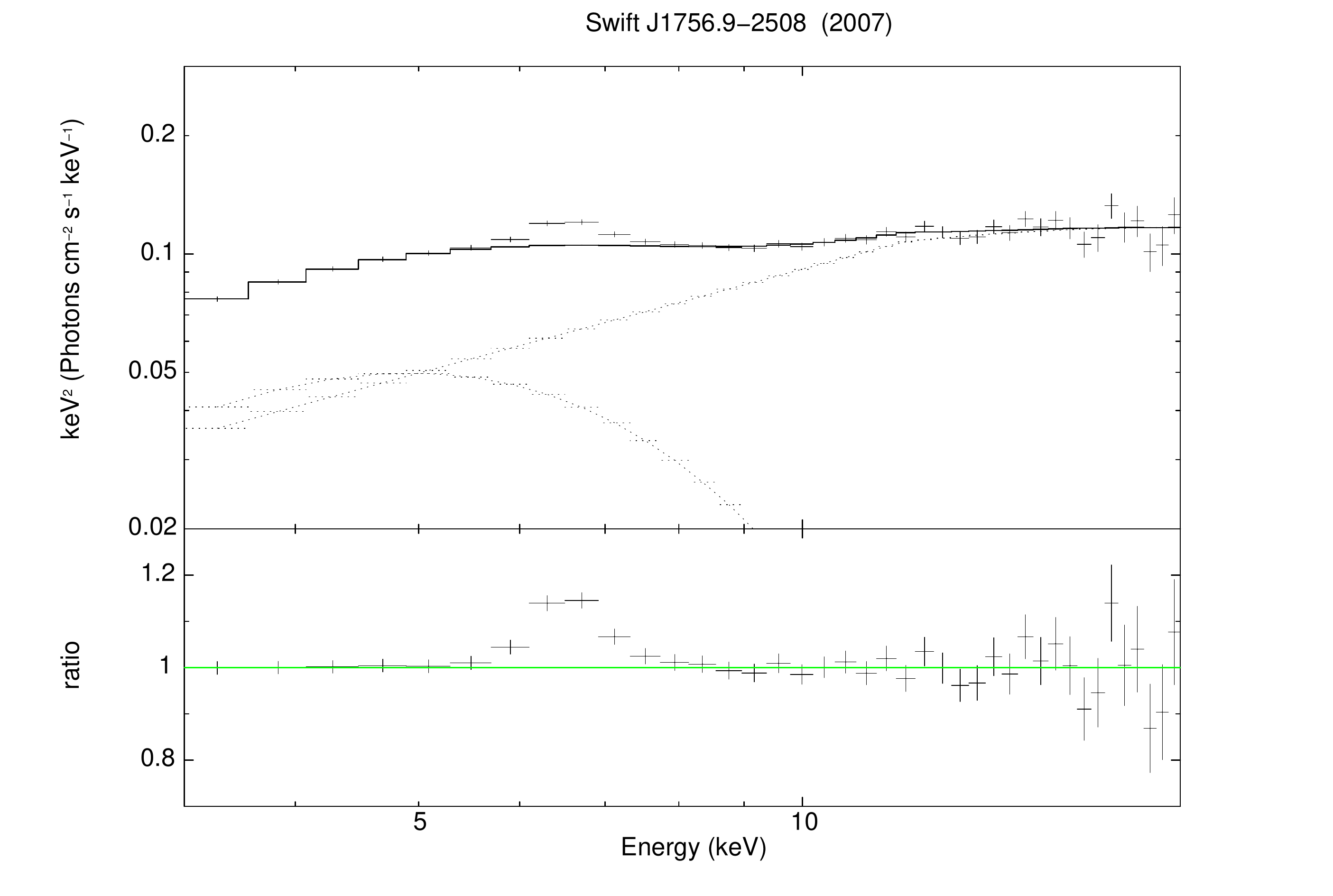}}
            \caption{Swift~J1756.9-2508 (2007): Unfolded spectrum. Energy and data-vs-model ratio plot, for only the continuum model. }
   \label{fig:swift_2007}
 \end{figure}
 
   \begin{figure}
       \resizebox{\hsize}{!}{\includegraphics[angle=-90,clip,trim=0 0 0 0,width=0.8\textwidth]{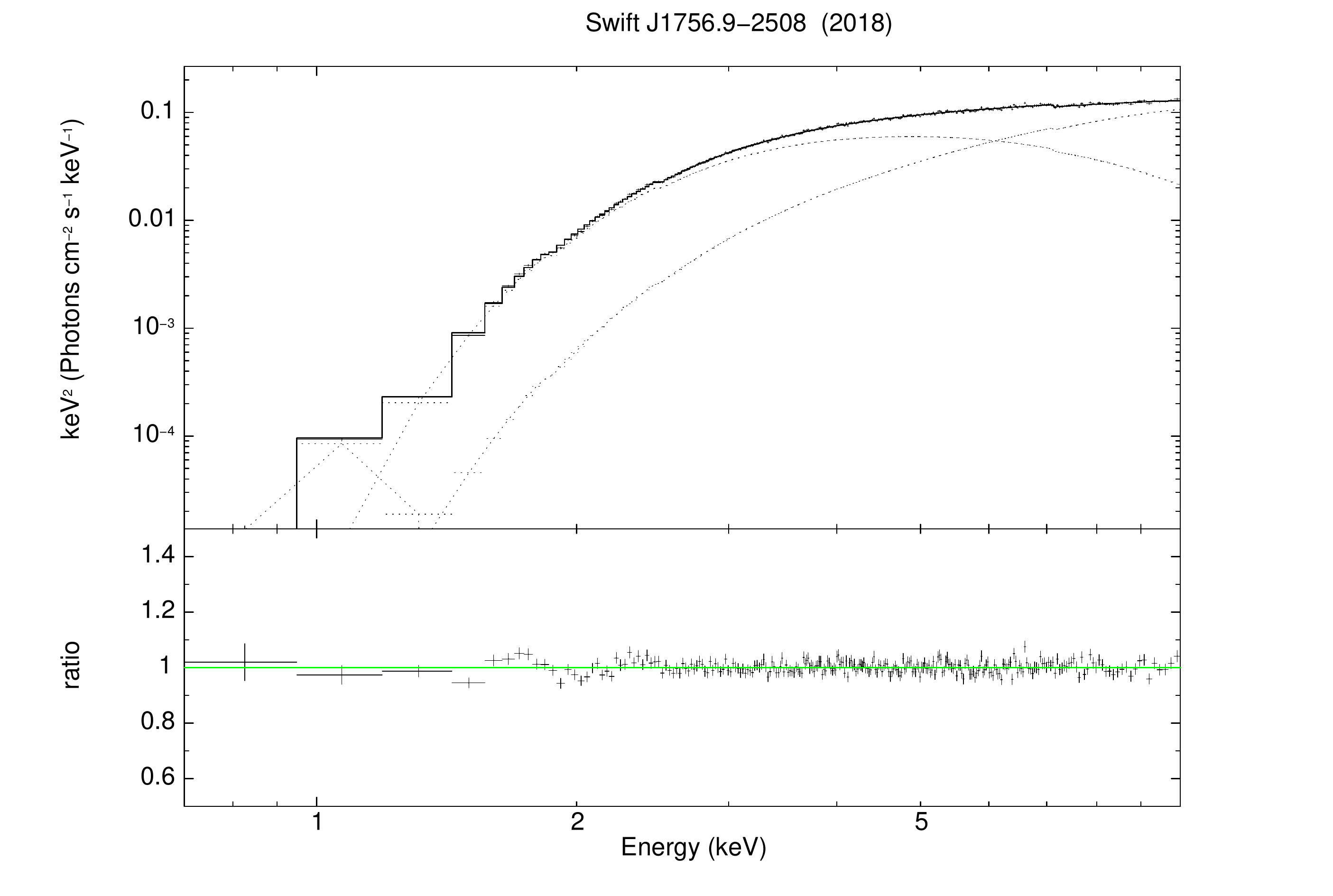}}
            \caption{Swift~J1756.9-2508 (2018): Unfolded spectrum. Energy and data-vs-model ratio plot, for only the continuum model. }
   \label{fig:swift}
 \end{figure}

  \begin{figure}
       \resizebox{\hsize}{!}{\includegraphics[angle=-90,clip,trim=0 0 0 0,width=0.8\textwidth]{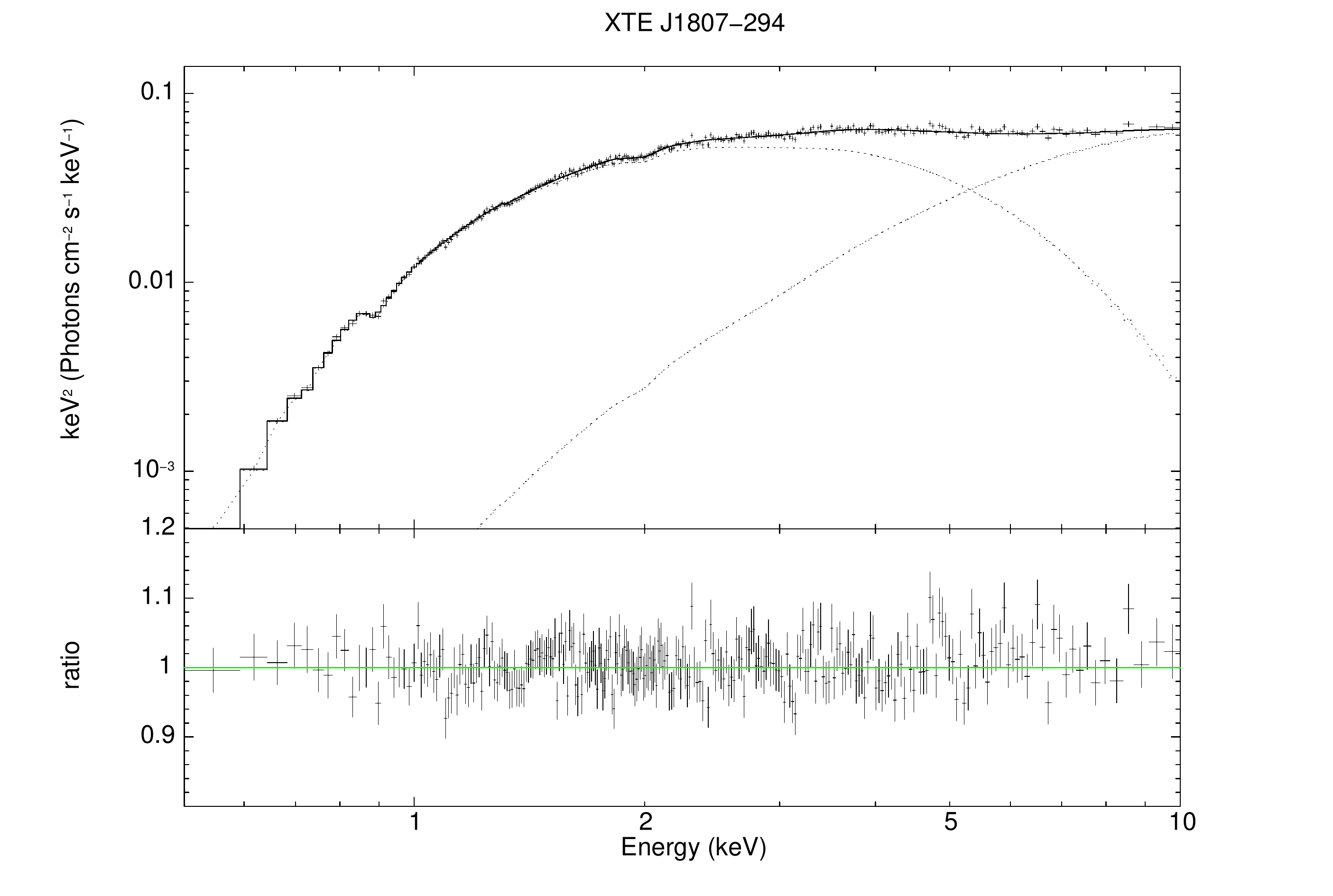}}
            \caption{XTE~J1807-294: Unfolded spectrum. Energy and data-vs-model ratio plot, for only the continuum model. }
   \label{fig:J1807}
 \end{figure}
 
  \begin{figure}
       \resizebox{\hsize}{!}{\includegraphics[angle=-90,clip,trim=0 0 0 0,width=0.8\textwidth]{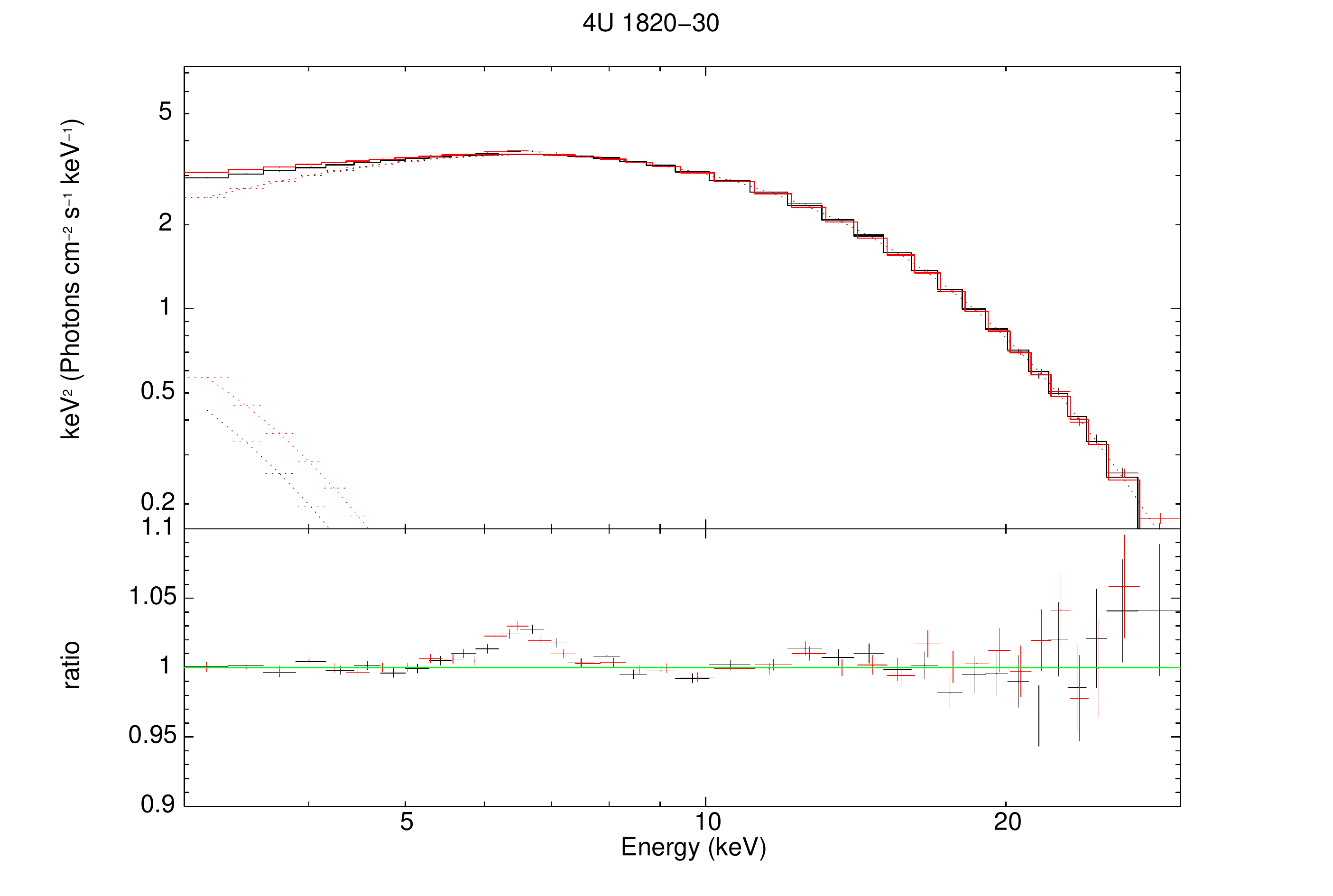}}
            \caption{4U~1820-30: Unfolded spectrum. Energy and data-vs-model ratio plot, for only the continuum model. }
   \label{fig:1820}
 \end{figure}
 
  \begin{figure}
       \resizebox{\hsize}{!}{\includegraphics[angle=-90,clip,trim=0 0 0 0,width=0.8\textwidth]{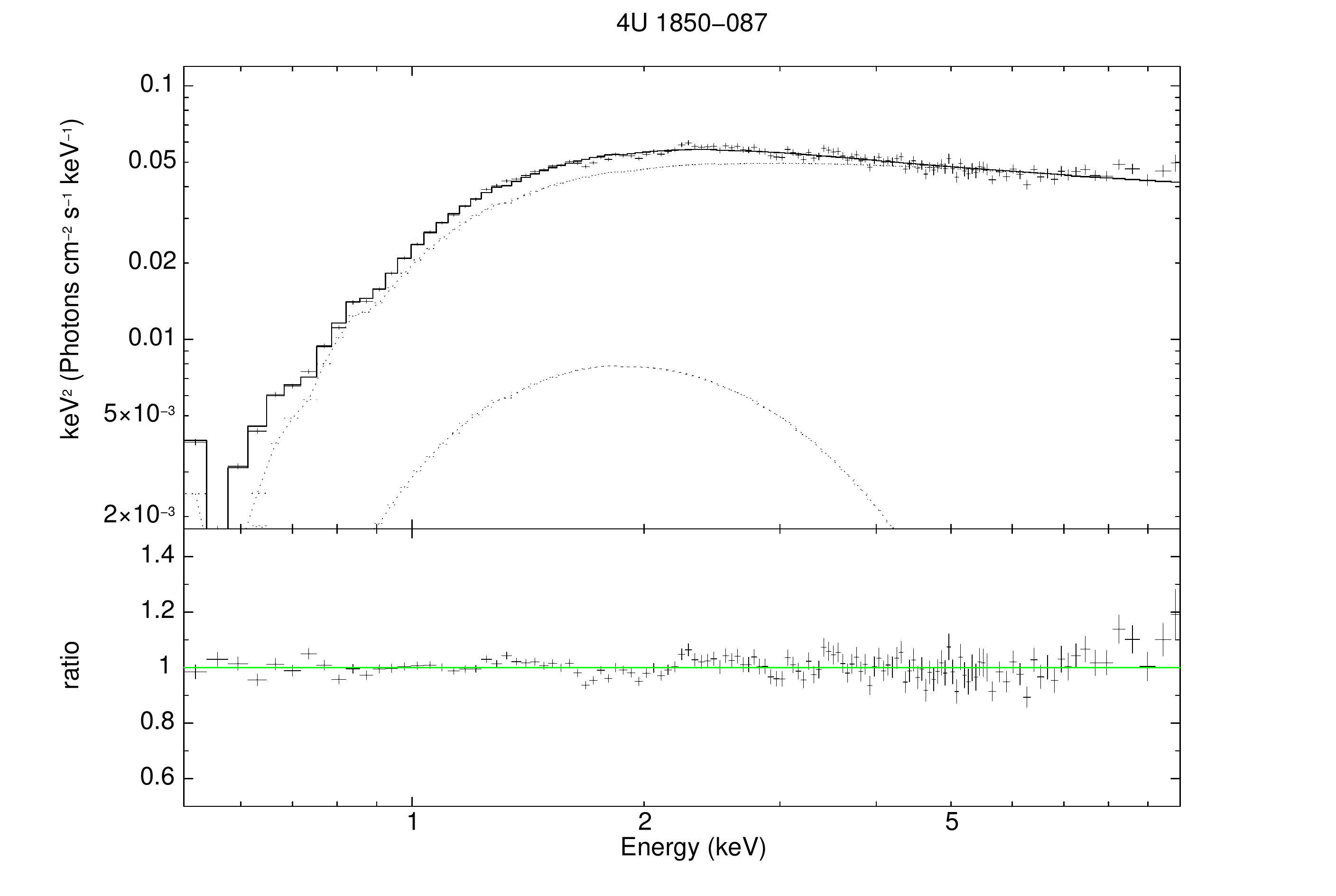}}
            \caption{4U~1850-087: Unfolded spectrum. Energy and data-vs-model ratio plot, for only the continuum model. }
   \label{fig:1850}
 \end{figure}
 
  \begin{figure}
       \resizebox{\hsize}{!}{\includegraphics[angle=-90,clip,trim=0 0 0 0,width=0.8\textwidth]{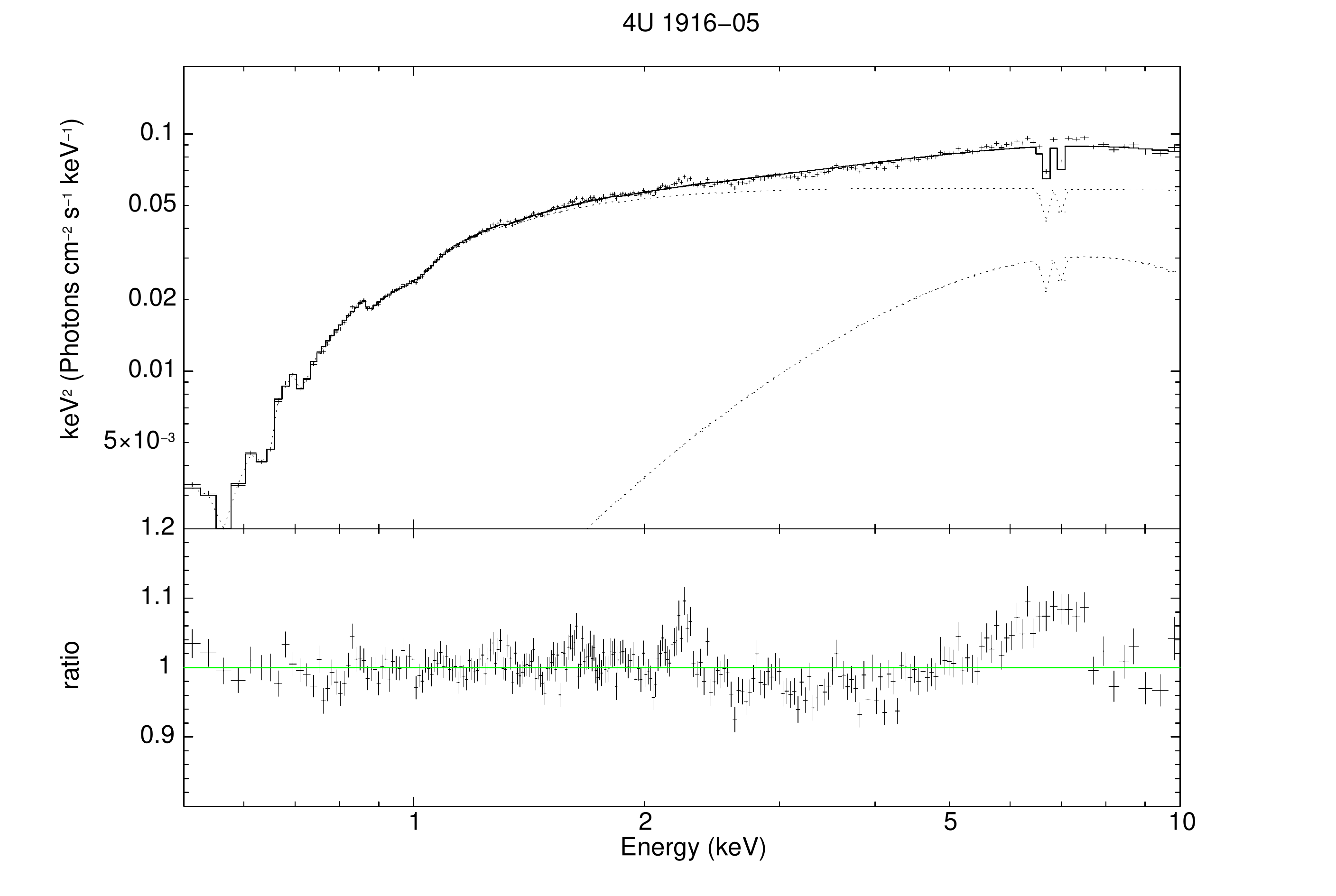}}
            \caption{4U~1916-05: Unfolded spectrum. Energy and data-vs-model ratio plot, for only the continuum model. }
   \label{fig:1916}
 \end{figure}
 
  \begin{figure}
       \resizebox{\hsize}{!}{\includegraphics[angle=-90,clip,trim=0 0 0 0,width=0.8\textwidth]{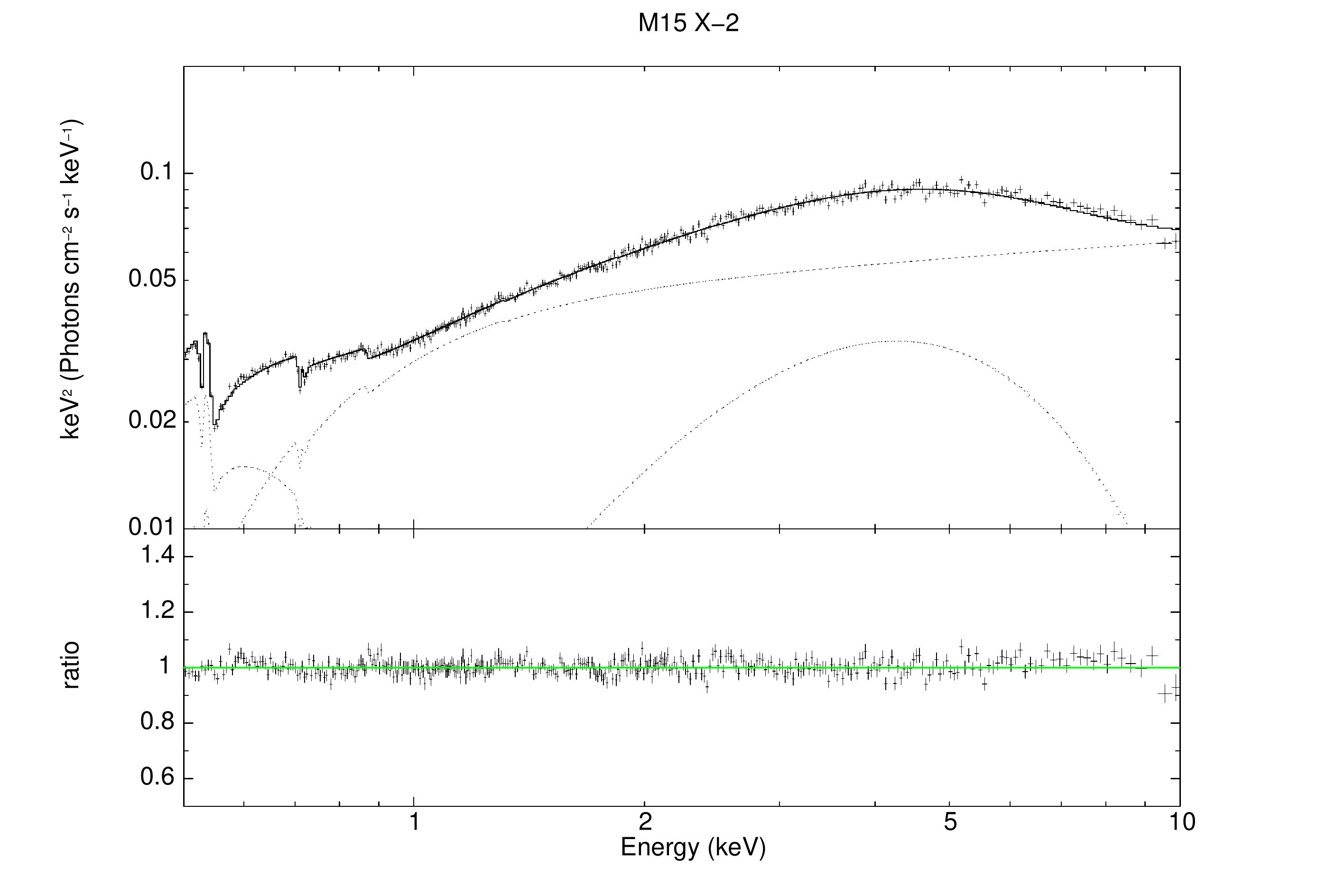}}
            \caption{M15~X-2: Unfolded spectrum. Energy and data-vs-model ratio plot, for only the continuum model. }
   \label{fig:M15}
 \end{figure}
 
   \begin{figure}
       \resizebox{\hsize}{!}{\includegraphics[angle=-90,clip,trim=0 0 0 0,width=0.8\textwidth]{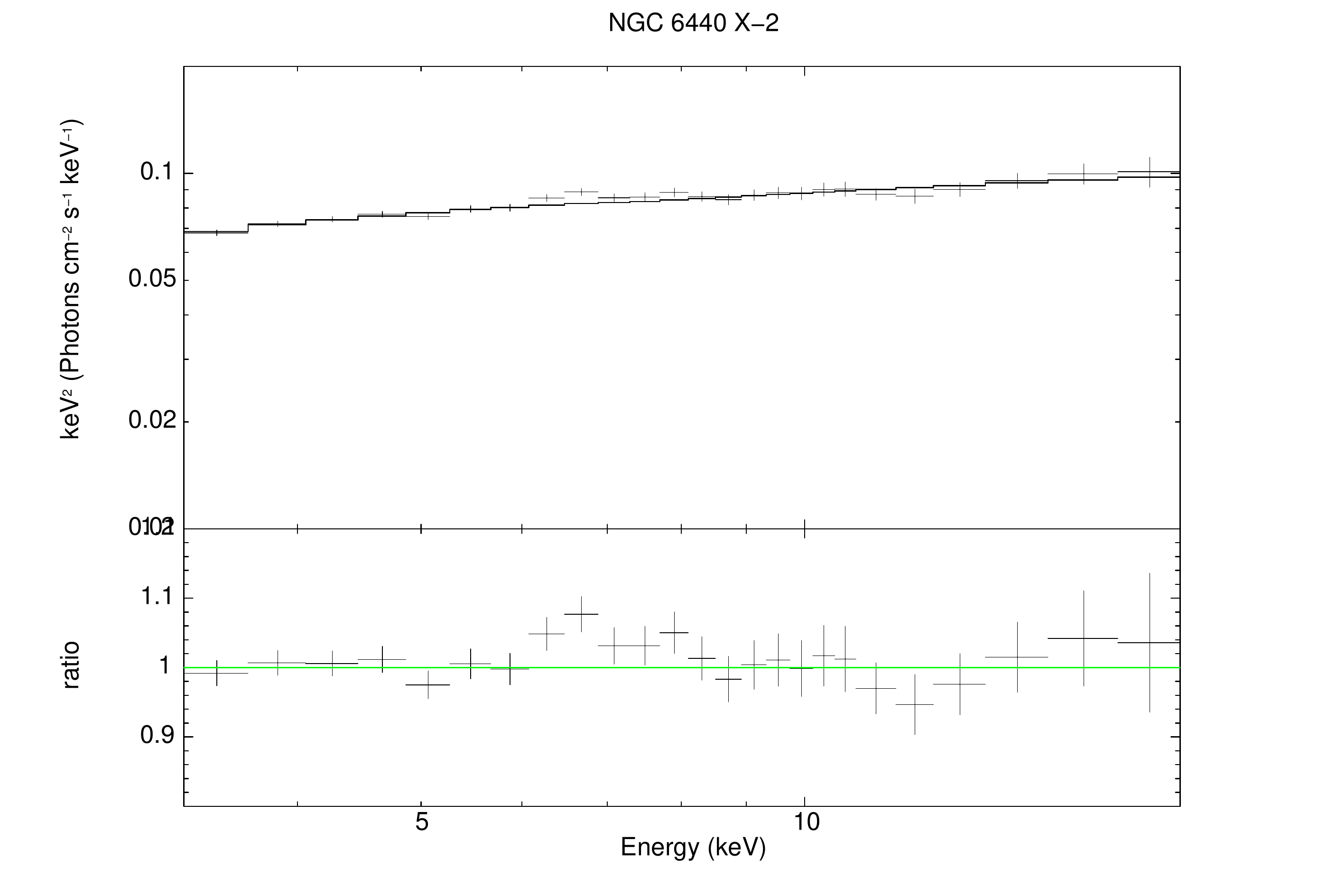}}
            \caption{NGC~6440~X-2: Unfolded spectrum. Energy and data-vs-model ratio plot, for only the continuum model. }
   \label{fig:NGC6440}
 \end{figure}

\bsp	
\label{lastpage}
\end{document}